\documentclass[twocolumn]{aastex6}
\usepackage{color,amsmath,textcomp,url,graphicx,subfigure,latexsym}
\usepackage{epsfig}
\usepackage{latexsym}% 
\usepackage{graphicx}%
\usepackage{amssymb}% 
\usepackage{url}
\usepackage{rotating}

\newcommand{\MJup}{$M_{\mathrm{Jup}}$}

\def\sun{\hbox{$\odot$}}
\def\earth{\hbox{$\oplus$}}

\def\farcs{\hbox{$.\!\!^{\prime\prime}$}}

\def\ref{\hangindent=15pt     % for references
        \hangafter=1}

\def\lesssim{\mathrel{\hbox{\rlap{\hbox{%
  \lower4pt\hbox{$\sim$}}}\hbox{$<$}}}}
\def\gtrsim{\mathrel{\hbox{\rlap{\hbox{%
  \lower4pt\hbox{$\sim$}}}\hbox{$>$}}}}

\def\farcs{\hbox{$.\!\!^{\prime\prime}$}}

\shorttitle{}
\shortauthors{Kellogg et al.}

\begin{document}

\title{A Statistical Survey of Peculiar L and T Dwarfs in SDSS, 2MASS, and WISE}
\author{Kendra Kellogg\altaffilmark{1,4}, Stanimir Metchev\altaffilmark{1,2}, Paulo A. Miles-P{\'a}ez\altaffilmark{1,3}, Megan E. Tannock\altaffilmark{1,4}}
\altaffiltext{1}{Department of Physics and Astronomy, Centre for Planetary and Space Exploration, The University of Western Ontario, 1151 Richmond St, London, ON N6A 3K7, Canada; kkellogg@uwo.ca, smetchev@uwo.ca}
\altaffiltext{2}{Department of Physics and Astronomy, Stony Brook University, Stony Brook, NY, 11794-3800, USA}
\altaffiltext{3}{Department of Astronomy and Steward Observatory, University of Arizona, 933 N Cherry Ave, Tucson, AZ 85719, USA}
\altaffiltext{4}{Visiting Astronomer at the Infrared Telescope Facility, which is operated by the University of Hawaii under contract NNH14CK55B with the National Aeronautics and Space Administration}

\begin{abstract}
We present the final results from a targeted search for brown dwarfs with unusual near-infrared colors. From a positional cross-match of SDSS, 2MASS and WISE, we have identified 144 candidate peculiar L and T dwarfs. Spectroscopy confirms that 20 of the objects are peculiar or are candidate binaries. Nine of the 420 objects in our sample are young ($\lesssim$200 Myr; 2.1\%) and another 8 (1.9\%) are unusually red with no signatures of youth. With a spectroscopic $J-K_s$ color of 2.58 $\pm$ 0.11~mag, one of the new objects, the L6 dwarf 2MASS J03530419+0418193, is among the reddest field dwarfs currently known and is one of the reddest objects with no signatures of youth known to date. We have also discovered another potentially very low gravity object, the L1 dwarf 2MASS J00133470+1109403, and independently identified the young L7 dwarf 2MASS J00440332+0228112, first reported by Schneider and collaborators. Our results confirm that signatures of low gravity are no longer discernible in low to moderate resolution spectra of objects older than $\sim$200 Myr. The 1.9\% of unusually red L dwarfs that do not show other signatures of youth could be slightly older, up to $\sim$400 Myr.  In this case a red $J-K_s$ color may be more diagnostic of moderate youth than individual spectral features.  However, its is also possible that these objects are relatively metal-rich, and so have an enhanced atmospheric dust content.

\end{abstract}


\maketitle

\section{Introduction} \label{sec:intro}

Ultra-cool dwarf atmospheres are complex: L-type dwarf atmospheres host a wide variety of atomic and molecular gases and mineral condensates, while the appearance of cooler T-type dwarfs is dominated by several molecular gas species and other more volatile elements. The change in spectral appearance from early- to late- L dwarfs follows an approximate monotonic trend with decreasing temperature throughout the spectral type sequence. However, across the L/T transition a drastic change in the appearance of the spectra takes place over only a narrow range of effective temperatures and luminosities \citep{kirkpatrick00,golimowski04,burgasser07}. 

A number of brown dwarf atmosphere models have been able to reproduce the observed characteristics of this transition by confining the condensate species to cloud layers (e.g., \citealp{ackerman01,marley02,tsuji02,burgasser02,burrows06}). As a brown dwarf cools and transitions from an L dwarf to a T dwarf, the optically thick clouds originally residing in the upper atmosphere sink and disappear below the photosphere. These clouds can vary in composition, height, structure and thickness. 

The clouds themselves are affected by a number of conditions. There have been numerous pieces of evidence to show that surface gravity is a contributing factor in the structure of clouds (e.g., \citealp{knapp04,cruz09,faherty12,faherty13,faherty16}). In young ultra-cool dwarfs, low surface gravity means that the clouds extend a greater range of altitudes in the atmosphere. This leads to redder near-infrared colors than their older counterparts at similar effective temperatures. There has also been evidence of unusually red brown dwarfs with high dust content that do not have signatures of youth \citep{looper08b,kirkpatrick10,liu16}. As there have not been many of these older red objects found, the cause of such dustiness is not well established. 

Settling the ambiguity in the underlying cause of unusually dusty atmospheres is undoubtedly of interest for understanding the evolution of substellar objects, and the processes that affect the sedimentation and/or condensation of atmospheric dust.  It is also crucial for revealing the ages and properties of directly imaged extrasolar planets, most of which exhibit spectral energy distribution (SED) characteristics of both youth and high dust content (e.g., \citealp{faherty13}). Because isolated brown dwarfs can be scrutinized much more readily than directly imaged extrasolar planets, we stand to potentially learn more about ultra-cool atmospheres from brown dwarfs than we can from exoplanets.

Observations of L+T binaries are also important for understanding the evolution of condensates and clouds in brown dwarf atmospheres across the L/T transition. Their coevality removes many of the uncertainties due to the distribution of initial conditions normally incorporated into evolutionary models. These types of systems also exhibit unusual near-infrared colors. In the $J$-band, the contributions from the L and T dwarf components are roughly equal --- the ``$J-$band bump" phenomenon in early T dwarfs (Tinney et al. 2003) --- while the contribution from the T dwarf is much less in the far-optical, resulting in slightly redder $z-J$ colors than for normal L dwarfs. The $J-K_s$ colors, however, are much bluer than normal L dwarfs because in the $K$-band, the contribution from the T dwarf is much fainter, while in the $J$-band the binary can be twice as bright as a single L dwarf. A number of such unresolved binaries have already been identified (e.g., \citealp{cruz04,burgasser10,bardalez14}) but the completeness of that set is unknown. The observations of L+T binaries have already shed light on several mysteries surrounding the L/T transition \citep{burgasser06,liu06,looper08b} and the role that clouds play in ultra-cool dwarf atmospheres (e.g. \citealp{ackerman01,burgasser02b,apai13}; Kellogg et al.\ 2017b, submitted). 

In view of our limited understanding of the evolution of substellar objects and the processes that affect condensation and sedimentation in the atmospheres of brown dwarfs, we carried out a dedicated search for L and T dwarfs with unusual optical/near-infrared colors. The goal was to substantially expand the sample of peculiar L and T dwarfs and L+T binaries in order to map the full range of their photospheric properties, and to better understand the evolution and content of L and T type atmospheres. We cross-correlated the SDSS, 2MASS and WISE catalogs to seek candidate peculiarly red brown dwarfs based solely on photometric criteria. From the first batch of candidates, presented in Kellogg et al.\ (\citeyear{kellogg15}, hereafter Paper 1 or P1), we discovered one of the brightest and least massive free-floating planetary-mass objects known to date, 2MASS J11193254--1137466 (\citealp{kellogg16}; TWA 42), which was recently resolved into a planetary-mass binary system where each object is $\sim$3\MJup\ \citep{best17b}. With this new survey, we determine the occurrence rate of various kinds of ultra-cool dwarfs by comparing our sample of peculiar L and T dwarfs to our full sample of ultra-cool dwarfs.

We discuss our candidate selection technique in $\S$\ref{sec:selection} and our follow-up spectroscopic observations in $\S$\ref{sec:obs}. We present our results and discuss the characteristics of all the objects we have identified as peculiar in $\S$\ref{sec:specclass}. In $\S$\ref{sec:stats} we discuss the totality of our results and put them into a brown dwarf evolutionary context and we present our conclusions in $\S$\ref{sec:concl}. 

\section{Candidate Selection} \label{sec:selection}
 
We implemented a photometric search for peculiar L and T dwarfs using combined optical (Sloan Digital Sky Survey; SDSS), near-infrared (2-Micron All-Sky Survey; 2MASS) and mid-infrared (Wide-Field Infrared Survey Explorer; WISE) fluxes. We applied joint positional and color criteria to the full SDSS DR9 and 2MASS point source catalogs to identify L and T dwarfs with unusual photometric colors. Our criteria included selecting objects with steep red optical slopes ($i-z >$ 1.5~mag; $z-J >$ 2.5~mag) and no counterparts in SDSS at wavelengths shorter than the $z$-band. We then cross-matched the results against the AllWISE catalog to confirm our ultra-cool dwarf candidates were detected in the mid-infrared where most brown dwarf energy distributions peak and had colors consistent with other ultra-cool dwarfs ($H-W2 >$ 1.2~mag). We identified sub-samples of potentially interesting candidates using criteria that selected objects with peculiar optical/near-infrared colors. Our full candidate selection process is detailed in P1 and we briefly discuss the various selection criteria in the following sections.  

\subsection{Candidate Ultra-cool Dwarfs} \label{subsec:bds}

In P1, we reported a sample of 314 objects that passed all of our selection criteria and visual verification. We had obtained spectra of a first set of 45 of these candidates and 5 of them turned out to be false positives, i.e. their spectra were not like those of ultra-cool dwarfs. The first 40 bonafide ultra-cool dwarfs were reported in P1. We reviewed the finder charts of the 5 false-positives and determined that these objects indeed did not look like the bonafide ultra-cool dwarfs (e.g. were more diffuse than point-like). After refining our visual verification, as informed by our re-analysis of the 5 false-positives, our total candidate L and T dwarf list was cut to 156 objects including 104 new candidates, the 40 candidates reported in P1, the original 5 false positives, and 7 new false positives. For this paper we will focus on the remaining 104 candidates that we verified to be ultra-cool dwarfs via spectroscopy and only briefly discuss the 12 total false positives in $\S$\ref{sec:specclass}. We also recovered 276 ultra-cool dwarfs that were previously known. We discuss these in $\S$\ref{sec:stats} where we analyze the statistics of the whole survey.

\subsection{Peculiarly Red Candidates} \label{subsec:redcand}

The priority of our survey was to identify peculiarly red ultra-cool dwarfs in the cross section of the SDSS, 2MASS and WISE catalogs. We designed an appropriate selection criterion from the sample of L and T dwarfs in the SpeX Prism Archive. We formed synthetic photometry from their spectra by convolving with the 2MASS filter transmission profiles and integrating over the filter bandpasses. We identified objects that had $J-K_s$ colors that were $>$2$\sigma$ redder than the median for the spectral type (red symbols in Figure~\ref{fig:ccdiag}a). The medians and standard deviations of the $J-K_s$ colors were taken from Faherty et al.\ (\citeyear{faherty09}; M7--M9 and T0--T8) and Faherty et al.\ (\citeyear{faherty13}; L0--L9). All of the red color outliers in the SpeX Prism Archive lie above the $z-J=-0.75(J-K_s)+3.8$~mag line in Figure~\ref{fig:ccdiag}a. We then applied this criterion to our 156 ultra-cool dwarf candidates and ended up with a sample of 88 peculiarly red candidates, 22 of which were already reported in P1. The synthetic colors of all of the candidates are presented in Figure~\ref{fig:ccdiag}b with different plotting colors used to represent objects with various spectroscopic peculiarities (discussed in $\S$\ref{sec:specclass}). Our selection criteria were based on the photometric colors of the candidates so there are a number of objects whose synthetic colors do not appear to pass the initial color-selection criteria (discussed more in $\S$\ref{sec:specclass}).

\subsection{Candidate T Dwarfs or L+T Binaries} \label{subsec:bincand}

\begin{figure*}
\centering 
\includegraphics[scale=0.45]{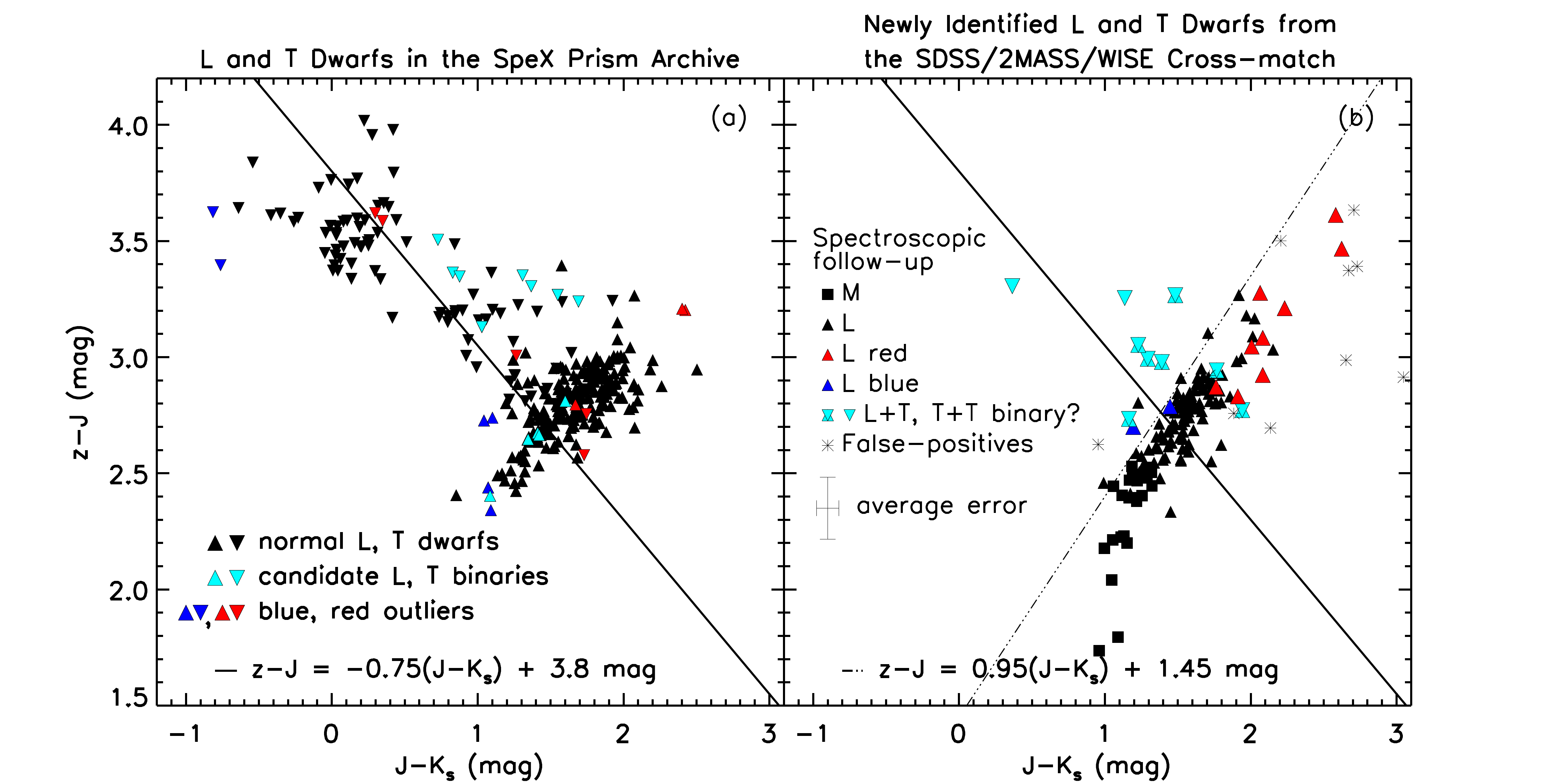}
\caption{\footnotesize (a) SDSS/2MASS synthetic color-color diagram of L and T dwarfs from the SpeX Prism Archive (upwards and downwards triangles, respectively). The $z-J$, and $J-K_s$ colors were formed synthetically from the SpeX spectra. Two-sigma red and blue photometric color outliers within each spectral type are indicated by red and blue symbols, respectively. The $z-J=-0.75(J-K_s)+3.8$~mag line was designed to select the red outliers based on their photometric SDSS/2MASS colors. (b) Color-color diagram of all of our L and T dwarf survey candidates with photometric colors redder than $z-J$ = 2.5 mag identified from the SDSS-2MASS-WISE cross-match. All symbols (squares - M dwarfs; upwards triangles - L dwarfs; downwards triangles - T dwarfs) represent the synthetic colors of the candidates from their spectra. The GNIRS spectra do not cover the entire $z$-band so for objects observed with GNIRS, the $z-J$ colors are their photometric colors. The black symbols are ``normal" objects and the red and blue symbols are objects that we have identified as peculiar or binary. Objects to the right of the $z-J=-0.75(J-K_s)+3.8$~mag line are candidate peculiarly red L and T dwarfs and objects to the left of the $z-J=0.95(J-K_s)+1.45$~mag line are candidate unresolved L+T binaries or T dwarfs. Both sets of peculiar objects were prioritized for spectroscopic follow-up. The photometric colors of the 12 false-positives ($\S$\ref{subsec:fp}) are shown by asterisks.}
\label{fig:ccdiag}
\end{figure*}

In addition to selecting unusually red objects, our prioritization criterion from $\S$\ref{subsec:redcand} also efficiently identified candidated unresolved L+T binary brown dwarfs. Figure~\ref{fig:ccdiag}b shows that they also stand out from the locus of objects on a $z-J$ vs $J-K_s$ diagram. Late-L and early-T dwarfs are similar in brightness in the $J$-band but are fainter in the $z$- and $K$-bands resulting in moderate or blue $J-K_s$ colors but red $z-J$ colors. To fully include all potential unresolved L+T binaries, we created a second independent criterion to select these. Any object that satisfied the criterion $z-J > 0.95(J-K_s)+1.45$~mag was either a candidate L+T binary or a candidate T dwarf as the latter also have the same red-optical and near-infrared colors. This criterion selected 13 objects that were candidate binary or T dwarfs. Twenty objects satisfied both selection criteria, i.e., they were red in $z-J$ but moderate in $J-K_s$ (top center of Fig.\ \ref{fig:ccdiag}b).

\section{Spectroscopic Observations and Data Reduction} \label{sec:obs}

We obtained near-infrared spectroscopic observations of the remaining 104 objects in our survey --- 66 peculiarly red, 13 candidate binary and 25 general ultra-cool dwarf candidates --- using the SpeX instrument \citep{rayner03} on the NASA Infrared Telescope Facility (IRTF) and the Gemini Near-Infrared Spectrograph (GNIRS) instrument \citep{elias06} on the Gemini North telescope. All reduction of the SpeX spectra was done in Interactive Data Language (IDL). The GNIRS spectra were reduced using the Gemini package version 1.13.1 in Image Reduction and Analysis Facility (IRAF; \citealp{cooke05}). 

\subsection{IRTF/SpeX} \label{subsec:spex}
We obtained the majority of our follow-up observations (91 of 104) with the SpeX spectrograph on the IRTF in prism mode (0.75--2.5~$\micron$; R$\sim$75--150), between 2014 October and 2016 April. Observing sequences and instrument settings were the same as those in P1. Table \ref{tab:spex} gives observation epochs and SpeX instrument settings for each science target. We reduced all the SpeX data in the same way as in P1. Figure \ref{fig:spectra} shows all reduced spectra in order of increasing spectral type (see $\S$~\ref{sec:specclass}) and within each spectral type in order of increasing RA.

\begin{deluxetable*}{lccccr}
\tabletypesize{\small}
\tablecolumns{6}
\tablewidth{0pt}
\tablecaption{IRTF/SpeX Observations
\label{tab:spex}}
\tablehead{
\colhead{Identifier} & \colhead{Date} & \colhead{2MASS J} & \colhead{Slit Width} & \colhead{Exposure} & \colhead{A0 Calibrator} \\
\colhead{(J2000)} & \colhead{(UT)} & \colhead{(mag)} & \colhead{(arcsec)} & \colhead{(min)} & \colhead{$ $} }
\startdata
\object{2MASS J00065552+0236376} & 2015 Oct 08 & 16.75 & 1.6 & 26 & BD+02 66\\
\object{2MASS J00062250+1300451} & 2015 Nov 30 & 16.96 & 1.6 & 24 & BD+12 5\\
\object{2MASS J00082822+3125581} & 2015 Jul 06 & 15.61 & 1.6 & 24 & HD 3925\\
\object{2MASS J00100480--0930519} & 2015 Jul 07 & 16.33 & 1.6 & 12 & HIP 115119\\
\object{2MASS J00132229--1143006} & 2014 Oct 12 & 16.35 & 0.8 & 40 & HIP 5899\\
\object{2MASS J00133470+1109403} & 2015 Sep 08 & 15.72 & 1.6 & 26 & BD+10 102\\
\object{2MASS J00150673+3006004} & 2014 Oct 12 & 16.10 & 0.8 & 20 & HIP 2969\\
\object{2MASS J00435012+0928429} & 2014 Oct 12 & 16.18 & 0.8 & 40 & HIP 10441\\
\object{2MASS J00440332+0228112\tablenotemark{a}} & 2015 Jul 01 & 17.00 & 1.6 & 38 & HD 9538\\
\object{2MASS J00452972+4237438} & 2015 Nov 30 & 17.06 & 1.6 & 60 & HIP 10559\\
\object{2MASS J00501561+1012431} & 2015 Jun 30 & 16.78 & 1.6 & 32 & HD 7353\\
\object{2MASS J00550564+0134365} & 2015 Jul 05 & 16.44 & 1.6 & 28 & HD 9538\\
\object{2MASS J01001471--0301494} & 2015 Jun 29 & 16.32 & 1.6 & 24 & HD 9716\\
\object{2MASS J01114355+2820024} & 2015 Jul 06 & 16.34 & 1.6 & 24 & HD 10681\\
\object{2MASS J01145788+4318561} & 2015 Jul 07 & 14.51 & 1.6 & 12 & HD 10773\\
\object{2MASS J01141304+4354287} & 2015 Nov 29 & 16.81 & 1.6 & 36 & HIP 13917\\
\object{2MASS J01165802+4333081} & 2015 Jul 07 & 16.88 & 1.6 & 36 & HD 10499\\
\object{2MASS J01183399+1810542} & 2015 Sep 08 & 15.72 & 1.6 & 34 & HD 10982\\
\object{2MASS J01194279+1122427} & 2014 Oct 12 & 15.97 & 0.8 & 20 & HIP 9965\\
\object{2MASS J01343635--0145444} & 2015 Nov 29 & 16.64 & 1.6 & 56 & HIP 13917\\
\object{2MASS J01341675--0546530} & 2015 Sep 08 & 16.17 & 1.6 & 34 & HD 7194\\
\object{2MASS J01352531+0205232} & 2015 Sep 08 & 16.62 & 1.6 & 34 & HD 7194\\
\object{2MASS J01392388--1845029} & 2015 Nov 30 & 16.55 & 1.6 & 60 & HIP 10185\\
\object{2MASS J01394906+3427226} & 2015 Oct 08 & 17.13 & 1.6 & 24 & HIP 10559\\
\object{2MASS J01414428+2227409} & 2015 Sep 08 & 16.81 & 1.6 & 50 & HD 14334\\
\object{2MASS J01442482-0430031} & 2015 Nov 29 & 17.25 & 1.6 & 60 & HIP 10512\\
\object{2MASS J01453520--0314117\tablenotemark{a}} & 2015 Oct 08 & 17.12 & 1.6 & 36 & HIP 10512\\
\object{2MASS J02151451+0453179} & 2015 Nov 30 & 16.60 & 1.6 & 24 & HIP 13917\\
\object{2MASS J02314893+4521059} & 2015 Nov 30 & 16.55 & 1.6 & 32 & HIP 15925\\
\object{2MASS J03315828+4130486} & 2015 Nov 30 & 16.86 & 1.6 & 32 & HIP 18769\\
\object{2MASS J03511847--1149326} & 2015 Feb 24 & 16.35 & 0.8 & 44 & HIP 19053\\
\object{2MASS J03530419+0418193} & 2014 Oct 12 & 16.47 & 0.8 & 20 & HD 29838\\
\object{2MASS J04214620--0025072} & 2015 Oct 08 & 16.34 & 1.6 & 20 & HIP 22435\\
\object{2MASS J04232191--0803051} & 2015 Nov 29 & 16.27 & 1.6 & 20 & HIP 22435 \\
\object{2MASS J04234652+0843211} & 2014 Dec 30 & 16.18 & 0.8 & 40 & HIP 22923\\
\object{2MASS J04510592+0014394} & 2015 Nov 30 & 16.78 & 1.6 & 40 & HIP 25121\\
\object{2MASS J07244848+2506143} & 2014 Dec 30 & 16.48 & 0.8 & 48 & HIP 38722\\
\object{2MASS J07552723+1138485} & 2015 Feb 25 & 17.26 & 0.8 & 40 & HIP 43018\\
\object{2MASS J08270185+4129191} & 2014 Dec 30 & 15.91 & 0.8 & 68 & HIP 41798\\
\object{2MASS J08443811+2226161} & 2014 Dec 30 & 16.80 & 0.8 & 36 & HIP 50459\\
\object{2MASS J09053247+1339138} & 2015 Feb 24 & 17.26 & 0.8 & 76 & HIP 48414\\
\object{2MASS J09083688+5526401} & 2015 Nov 29 & 16.46 & 1.6 & 24 & HIP 50459\\
\object{2MASS J09194512+5135149} & 2015 Feb 25 & 16.72 & 0.8 & 40 & HIP 53735\\
\object{2MASS J09325053+1836485} & 2015 Nov 29 & 17.46 & 1.6 & 60 & HIP 50459\\
\object{2MASS J09393078+0653098\tablenotemark{b}} & 2015 Nov 30 & 16.78 & 1.6 & 44 & HIP 45167\\
\object{2MASS J09481259+5300387} & 2015 Nov 29 & 15.59 & 1.6 & 16 & HIP 53735\\
\object{2MASS J10271549+5445175} & 2016 Apr 24 & 16.15 & 1.6 & 26 & HIP 53735\\
\object{2MASS J10551343+2504028} & 2016 Apr 17 & 17.06 & 1.6 & 50 & HIP 55627\\
\object{2MASS J10592523+5659596} & 2016 Apr 24 & 15.51 & 1.6 & 34 & HIP 56147\\
\object{2MASS J11060459--1907025} & 2016 Apr 17 & 16.76 & 1.6 & 30 & HIP 56746\\
\object{2MASS J11213919--1053269} & 2016 Apr 24 & 16.44 & 1.6 & 30 & HD 97516\\
\object{2MASS J11220855+0343193} & 2016 Apr 17 & 16.65 & 1.6 & 10 & HIP 54849\\
\object{2MASS J11285958+5110202} & 2016 Apr 24 & 16.19 & 1.6 & 26 & HIP 52478\\
\object{2MASS J11282763+5934003} & 2016 Apr 24 & 16.37 & 1.6 & 36 & HD 97516\\
\object{2MASS J12023885+5345384} & 2016 Apr 24 & 17.56 & 1.6 & 68 & HD 108346\\
\object{2MASS J12232570+0448277\tablenotemark{b}} & 2016 Apr 16 & 16.33 & 1.6 & 18 & HIP 62745\\
\object{2MASS J12352675+4124310} & 2016 Apr 16 & 16.71 & 1.6 & 34 & HIP 65280\\
\object{2MASS J12453705+4028456} & 2016 Apr 17 & 16.75 & 1.6 & 26 & HIP 65280\\
\object{2MASS J12492272+0310255} & 2016 Apr 17 & 16.36 & 1.6 & 18 & HIP 62745\\
\object{2MASS J13042886-0032410} & 2016 Apr 16 & 17.03 & 1.6 & 26 & HIP 65599\\
\object{2MASS J13064517+4548552} & 2016 Apr 24 & 17.01 & 1.6 & 52 & HD 116405\\
\object{2MASS J13170488+3447513} & 2016 Apr 17 & 16.50 & 1.6 & 34 & HIP 61534\\
\object{2MASS J13184567+3626138\tablenotemark{a}} & 2016 Apr 17 & 17.21 & 1.6 & 34 & HIP 65280\\
\object{2MASS J13264464+3627407} & 2016 Apr 17 & 16.44 & 1.6 & 34 & HIP 65280\\
\object{2MASS J13451417+4757231} & 2016 Apr 24 & 16.50 & 1.6 & 26 & HIP 68767\\
\object{2MASS J14124574+3403074} & 2016 Apr 17 & 16.55 & 1.6 & 34 & HIP 71172\\
\object{2MASS J14154242+2635040} & 2016 Apr 24 & 16.37 & 1.6 & 34 & HIP 77111\\
\object{2MASS J14313545--0313117} & 2016 Apr 17 & 16.09 & 1.6 & 26 & HIP 73200\\
\object{2MASS J14554511+3843329} & 2016 Apr 24 & 16.70 & 1.6 & 34 & HIP 77111\\
\object{2MASS J15102256--1147125} & 2016 Apr 16 & 15.66 & 1.6 & 10 & HIP 78436\\
\object{2MASS J15163838+3333576} & 2016 Apr 16 & 16.79 & 1.6 & 42 & HIP 77111\\
\object{2MASS J15442544+0750572} & 2016 Apr 16 & 16.75 & 1.6 & 34 & HIP 79332\\
\object{2MASS J15500191+4500451} & 2016 Apr 17 & 17.33 & 1.6 & 66 & HD 141930\\
\object{2MASS J15525579+1123523} & 2016 Apr 17 & 15.92 & 1.6 & 18 & HIP 79332\\
\object{2MASS J15543602+2724487} & 2015 Sep 08 & 16.19 & 1.6 & 22 & HIP 77111\\
\object{2MASS J15565004+1449081} & 2015 Sep 08 & 17.31 & 1.6 & 38 & HIP 77111\\
\object{2MASS J16123860+3126489} & 2016 Apr 24 & 16.64 & 1.6 & 34 & BD+34 2755\\
\object{2MASS J17120142+3108217} & 2015 Jul 01 & 16.18 & 1.6 & 40 & HD 161259\\
\object{2MASS J17153111+1054108} & 2015 Jul 05 & 17.11 & 1.6 & 12 & HD 161259\\
\object{2MASS J17440969+5135032} & 2015 Jul 07 & 16.92 & 1.6 & 16 & HIP 82884\\
\object{2MASS J17570962+4325139} & 2015 Sep 14 & 16.73 & 1.6 & 50 & HD 170560\\
\object{2MASS J21123034+0758505} & 2015 Jul 06 & 16.26 & 1.6 & 32 & HD 207073\\
\object{2MASS J22035781+0713492} & 2015 Jun 19 & 16.68 & 1.6 & 28 & HIP 116886\\
\object{2MASS J22191282+1113405\tablenotemark{b}} & 2015 Oct 08 & 16.74 & 1.6 & 20 & HIP 109452\\
\object{2MASS J22295358+1556180} & 2015 Jul 01 & 16.46 & 1.6 & 32 & HD 116886\\
\object{2MASS J22355244+0418563} & 2014 Oct 12 & 15.37 & 0.8 & 20 & HIP 116886\\
\object{2MASS J22545900--0330590} & 2015 Nov 29 & 16.84 & 1.6 & 40 & HIP 116886\\
\object{2MASS J22582325+2906484} & 2015 Nov 29 & 16.77 & 1.6 & 36 & HIP 116886\\
\object{2MASS J23004298+0200145} & 2014 Oct 12 & 16.40 & 0.8 & 40 & HIP 116886\\
\object{2MASS J23053808+0524070} & 2014 Oct 12 & 16.43 & 0.8 & 40 & HIP 116886\\
\object{2MASS J23313131+2041273} & 2015 Jul 05 & 16.06 & 1.6 & 40 & HD 3347\\
\enddata
\tablenotetext{a}{\footnotesize Independently reported by \cite{schneider17}.}
\tablenotetext{b}{\footnotesize Independently reported by \cite{best17}.}
\end{deluxetable*}

\subsection{Gemini/GNIRS} \label{subsec:gnirs}
We followed-up the remaining 13 objects in our candidate list using GNIRS on Gemini North (0.9--2.5~$\micron$). We observed these objects in queue mode between 2015 October and 2017 May. We took the observations in cross-dispersed mode with the short-blue camera, 32 l/mm grating and $1\farcs0 \times 7\farcs0$ slit resulting in a resolution of R$\sim$500. We used a standard A-B-B-A nodding sequence along the slit to record object and sky spectra. Individual exposure times were 120s per pointing. Standard stars were used for flux calibration and telluric correction. Flat-field and argon lamps were taken immediately after each set of target and standard star observations for use in instrumental calibrations. Table \ref{tab:gnirs} gives Gemini/GNIRS observation epochs for each science target.

We reduced the cross-dispersed spectra by straightening the traces, rectifying them to the vertical, and then wavelength calibrating before extracting. We extracted the spectra using the variance weighted sum of the flux within the aperture with the aperture radius equal to the PSF radius (usually $\sim$4 pixels = 0$\farcs$60). We modeled a local background using a linear fit to a specified background region (usually $\sim$8 pixels = 1$\farcs$2 wide on either side of the PSF $\sim$2 pixels = 0$\farcs$30 away from the PSF) and subtracted it from the spectra which we subsequently extracted. Each set of extracted spectra were median-combined, corrected for telluric absorption and flux-calibrated with their associated A0 calibration star. We median combined all calibrated sets of observing sequences to produce a final spectrum. The reduced spectra were smoothed, using the IDL interpolation algorithm with a least squares quadratic fit, to the same resolution as the SpeX standards for comparison in $\S$\ref{sec:specclass}. The reduced Gemini/GNIRs spectra are included in Figure~\ref{fig:spectra}, where they are shown prior to smoothing.

\begin{deluxetable*}{lcccr}
\tabletypesize{\small}
\tablecolumns{5}
\tablewidth{0pt}
\tablecaption{Gemini/GNIRS Observations
\label{tab:gnirs}}
\tablehead{
\colhead{Identifier} & \colhead{Date} & \colhead{2MASS J} & \colhead{Exposure} & \colhead{A0 Calibrator} \\
\colhead{(J2000)} & \colhead{(UT)} & \colhead{(mag)} & \colhead{(min)} & \colhead{$ $} }
\startdata
\object{2MASS J01412651+1001339} & 2015 Oct 21 & 17.05 & 44 & HIP 7353\\
\object{2MASS J02022917+2305141\tablenotemark{a}} & 2015 Nov 04 & 17.22 & 48 & HD 9071\\
\object{2MASS J03302948+3910242} & 2016 Jan 05 & 17.12 & 60 & HIP 18769\\
\object{2MASS J09240328+3653444} & 2015 Nov 04 & 17.09 & 64 & HIP 41798\\
\object{2MASS J10265851+2515262} & 2016 Jan 25 & 17.36 & 56 & HIP 50459\\
\object{2MASS J10524963+1858151} & 2016 Jan 25 & 17.30 & 56 & HIP 56736\\
\object{2MASS J12260640+1756293} & 2015 Apr 02 & 16.85 & 40 & HIP 56736\\
\object{2MASS J14193789+3333326} & 2017 Feb 01 & 16.30 & 120 & HIP 68767\\
\object{2MASS J15025475+5044252} & 2017 Apr 12 & 16.16 & 48 & HIP 67848\\
\object{2MASS J15202471+2203340} & 2017 May 23 & 16.67 & 124 & HIP 68767\\
\object{2MASS J15552840+5918155} & 2017 Apr 15 & 15.96 & 120 & HIP 78017\\
\object{2MASS J16194822--0425366} & 2017 Apr 18 & 16.57 & 68 & HIP 81584\\
\object{2MASS J17164469+2302220} & 2015 Apr 02 & 17.02 & 56 & HIP 79102\\
\enddata
\tablenotetext{a}{\footnotesize Independently reported by \cite{schneider17}.}
\end{deluxetable*}

\begin{figure*}[t!]
\centering
\begin{subfigure}
  \centering
  \includegraphics[width=0.45\linewidth]{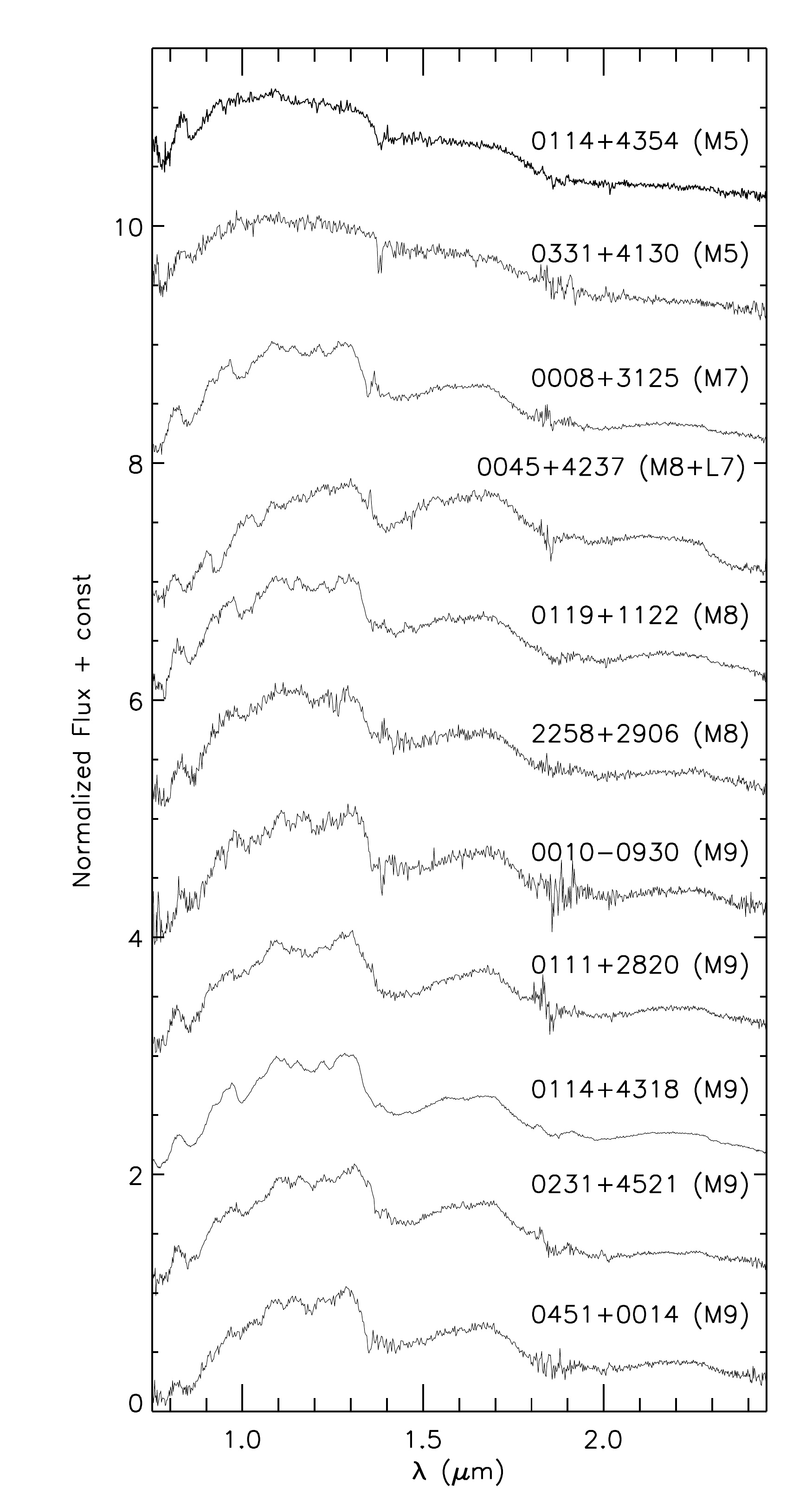}
\end{subfigure}%
\begin{subfigure}
  \centering
  \includegraphics[width=0.45\linewidth]{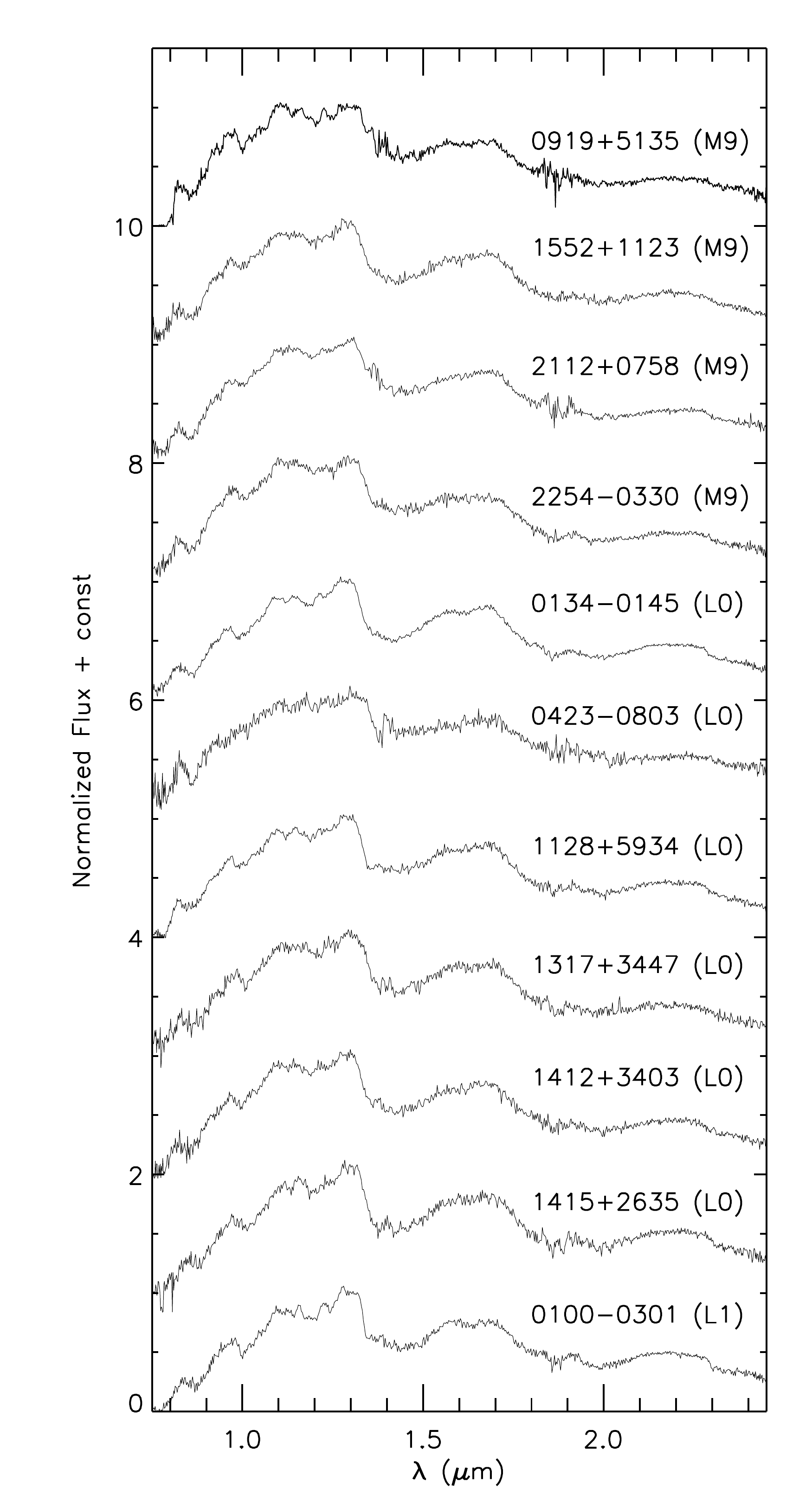}
\end{subfigure}
\caption{SpeX (0.75--2.5~$\micron$; $R \sim$75--150) and GNIRS (0.9--2.5~$\micron$; $R \sim$500) spectra of of the 104 newly discovered ultra-cool dwarfs in order of spectral type. Spectral types are given in parentheses.}
\label{fig:spectra}
\end{figure*}

\renewcommand{\thefigure}{\arabic{figure} (cont)}

\begin{figure*}[t!]
%\ContinuedFloat
\setcounter{figure}{1}
\centering
\begin{subfigure}
  \centering
  \includegraphics[width=0.45\linewidth]{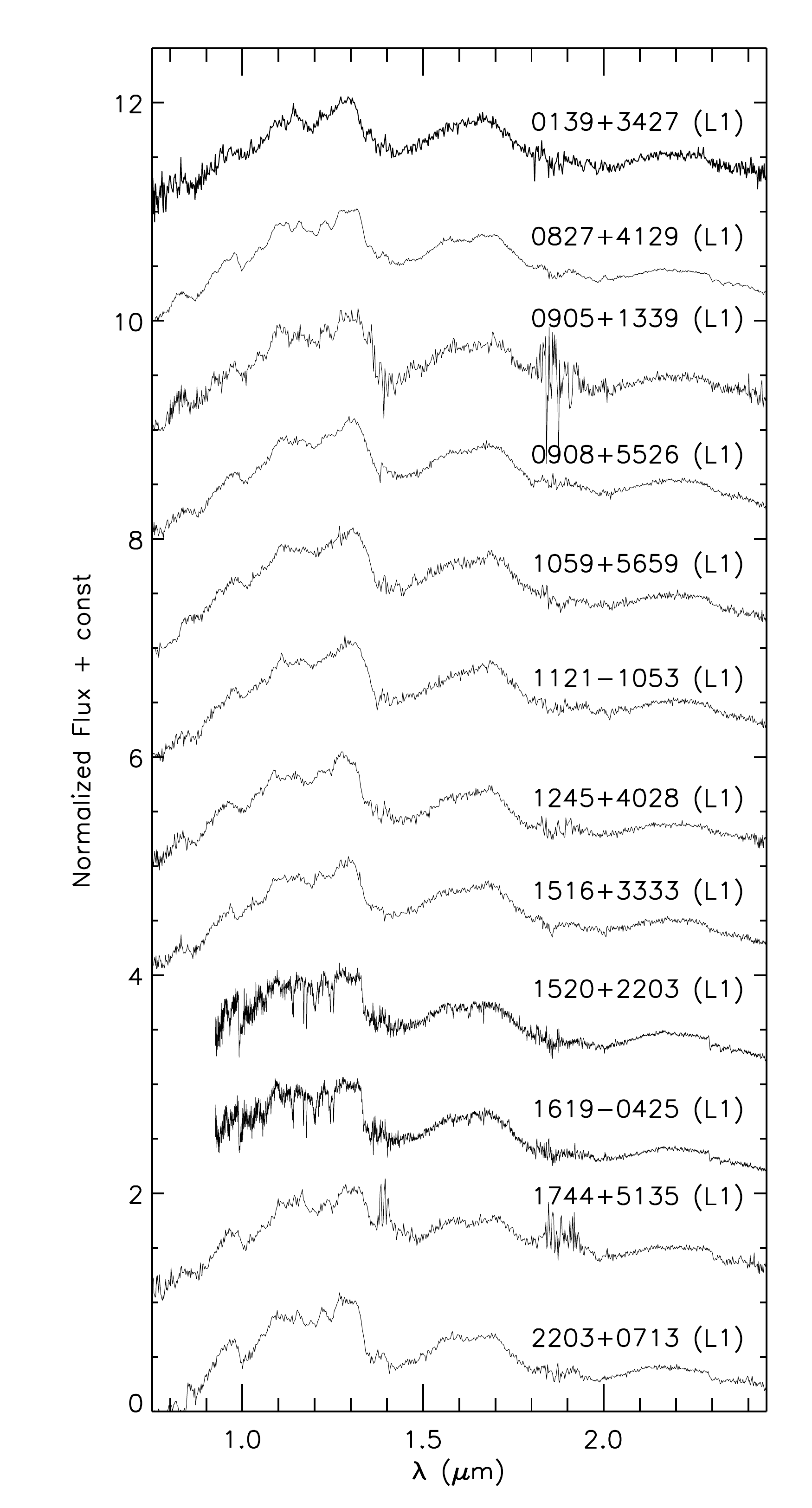}
\end{subfigure}%
\begin{subfigure}
  \centering
  \includegraphics[width=0.45\linewidth]{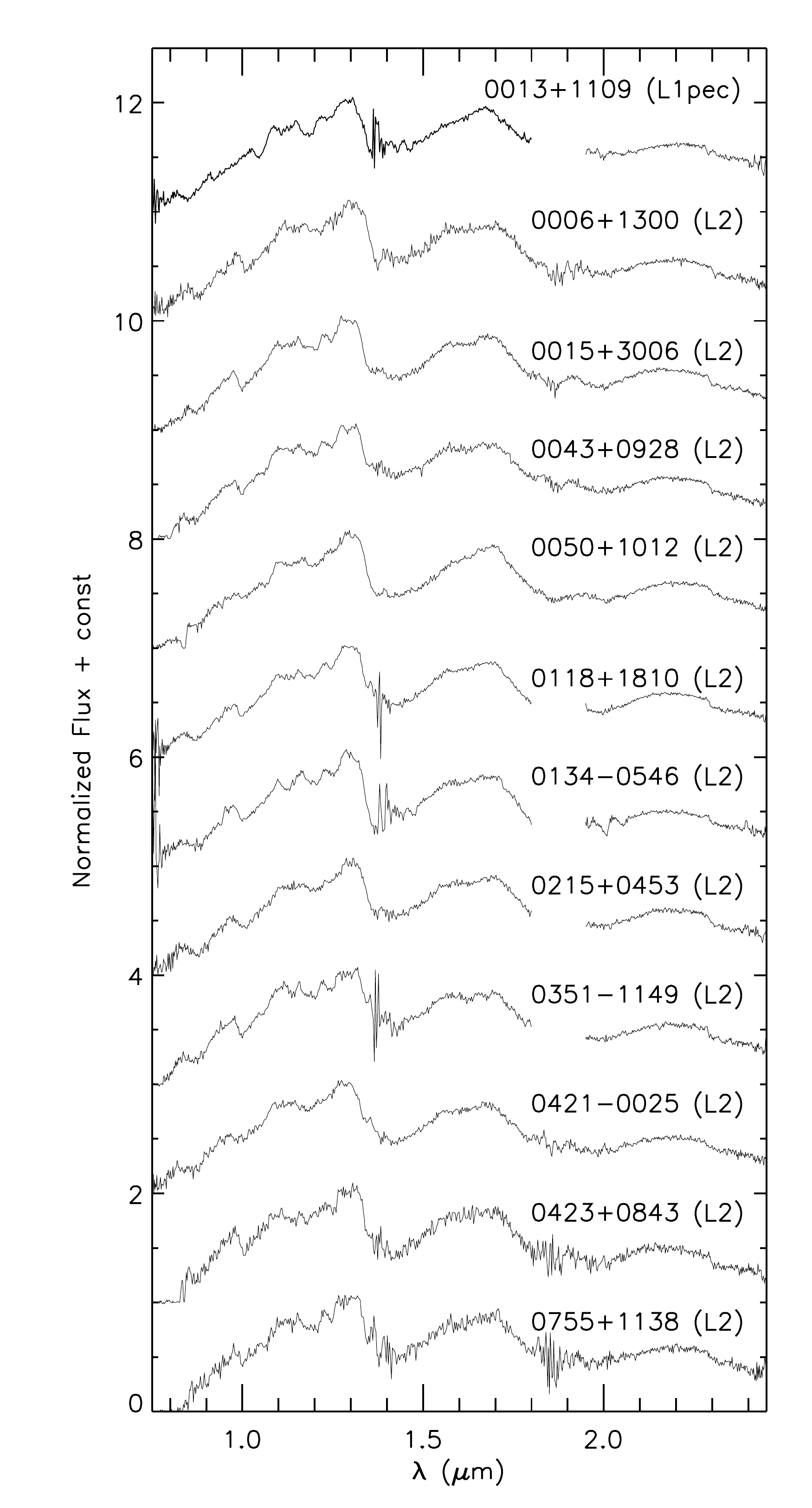}
\end{subfigure}
\caption{}
\end{figure*}

\begin{figure*}[t!]
%\ContinuedFloat
\setcounter{figure}{1}
\centering
\begin{subfigure}
  \centering
  \includegraphics[width=0.45\linewidth]{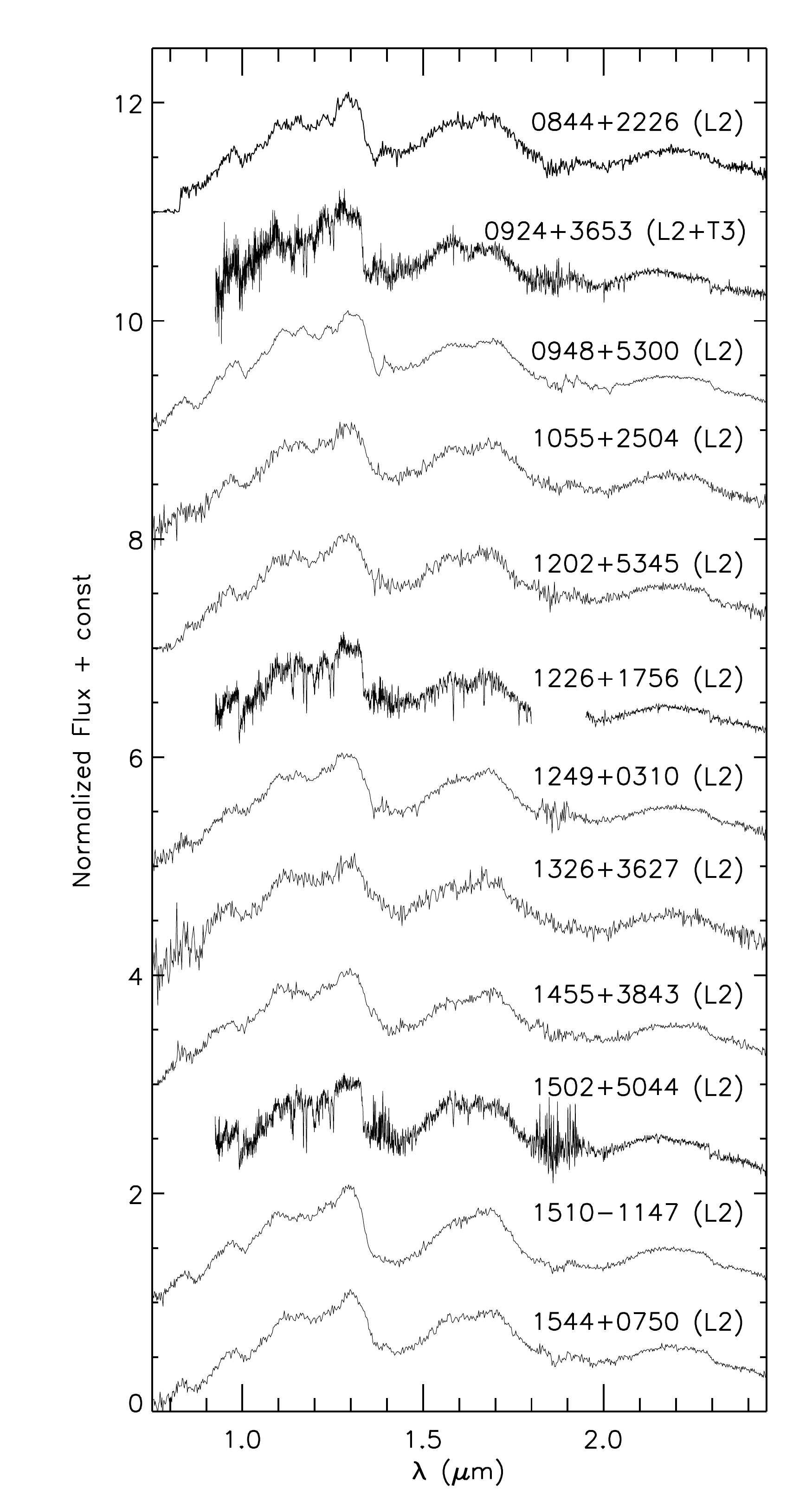}
\end{subfigure}%
\begin{subfigure}
  \centering
  \includegraphics[width=0.45\linewidth]{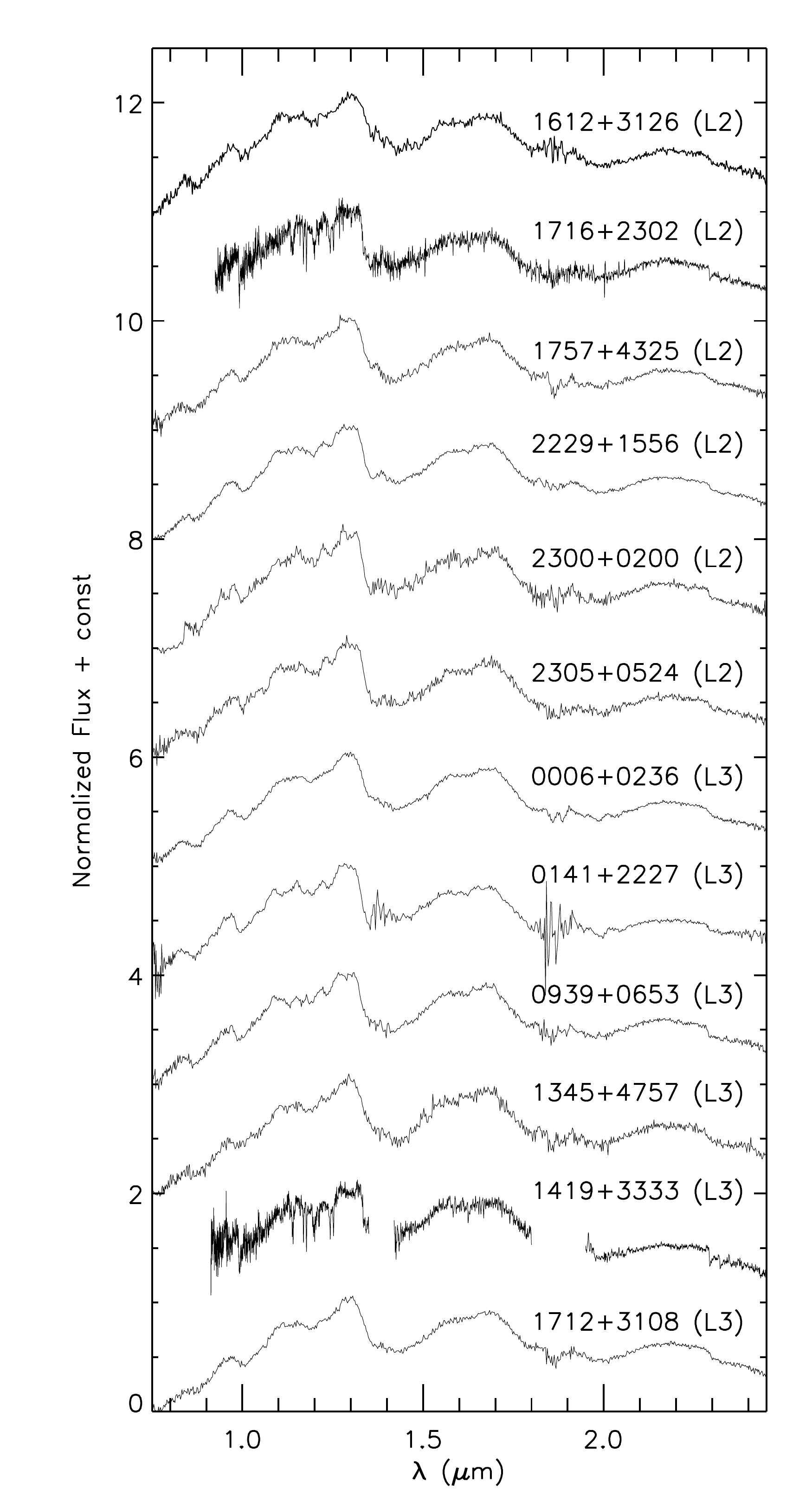}
\end{subfigure}
\caption{}
\end{figure*}

\begin{figure*}[t!]
%\ContinuedFloat
\setcounter{figure}{1}
\centering
\begin{subfigure}
  \centering
  \includegraphics[width=0.45\linewidth]{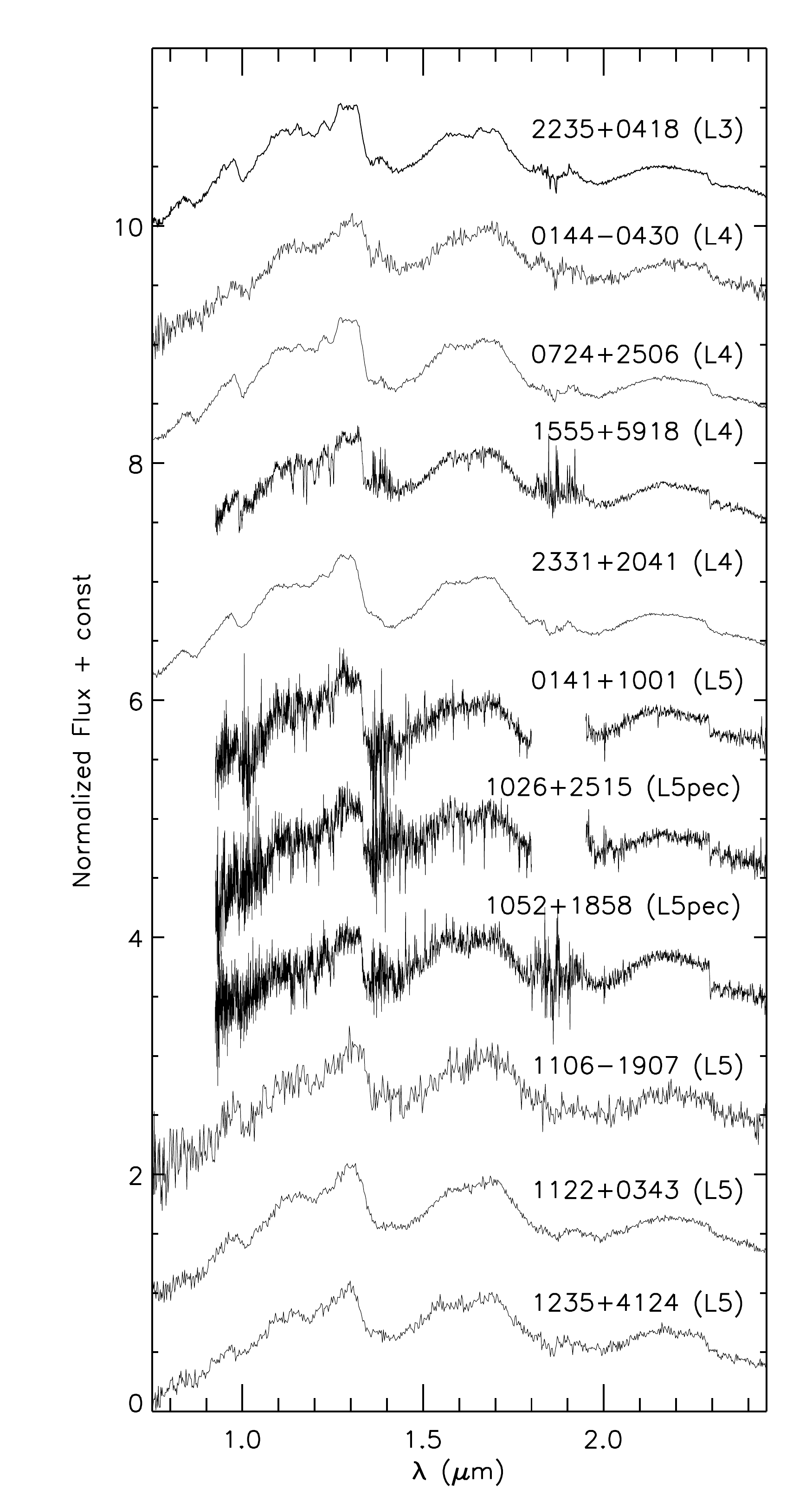}
\end{subfigure}%
\begin{subfigure}
  \centering
  \includegraphics[width=0.45\linewidth]{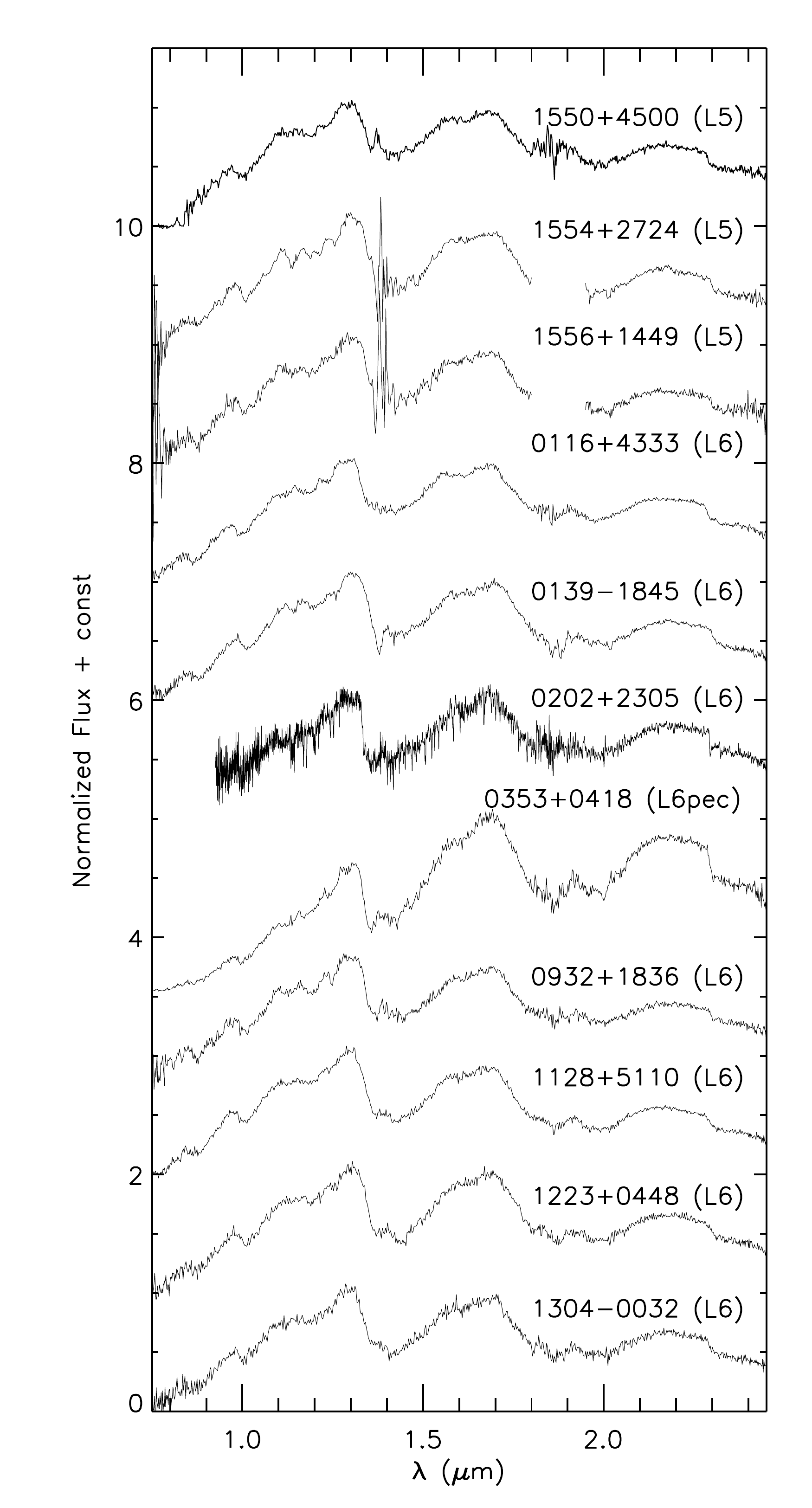}
\end{subfigure}
\caption{}
\end{figure*}

\begin{figure*}[t!]
%\ContinuedFloat
\setcounter{figure}{1}
\centering
\begin{subfigure}
  \centering
  \includegraphics[width=0.45\linewidth]{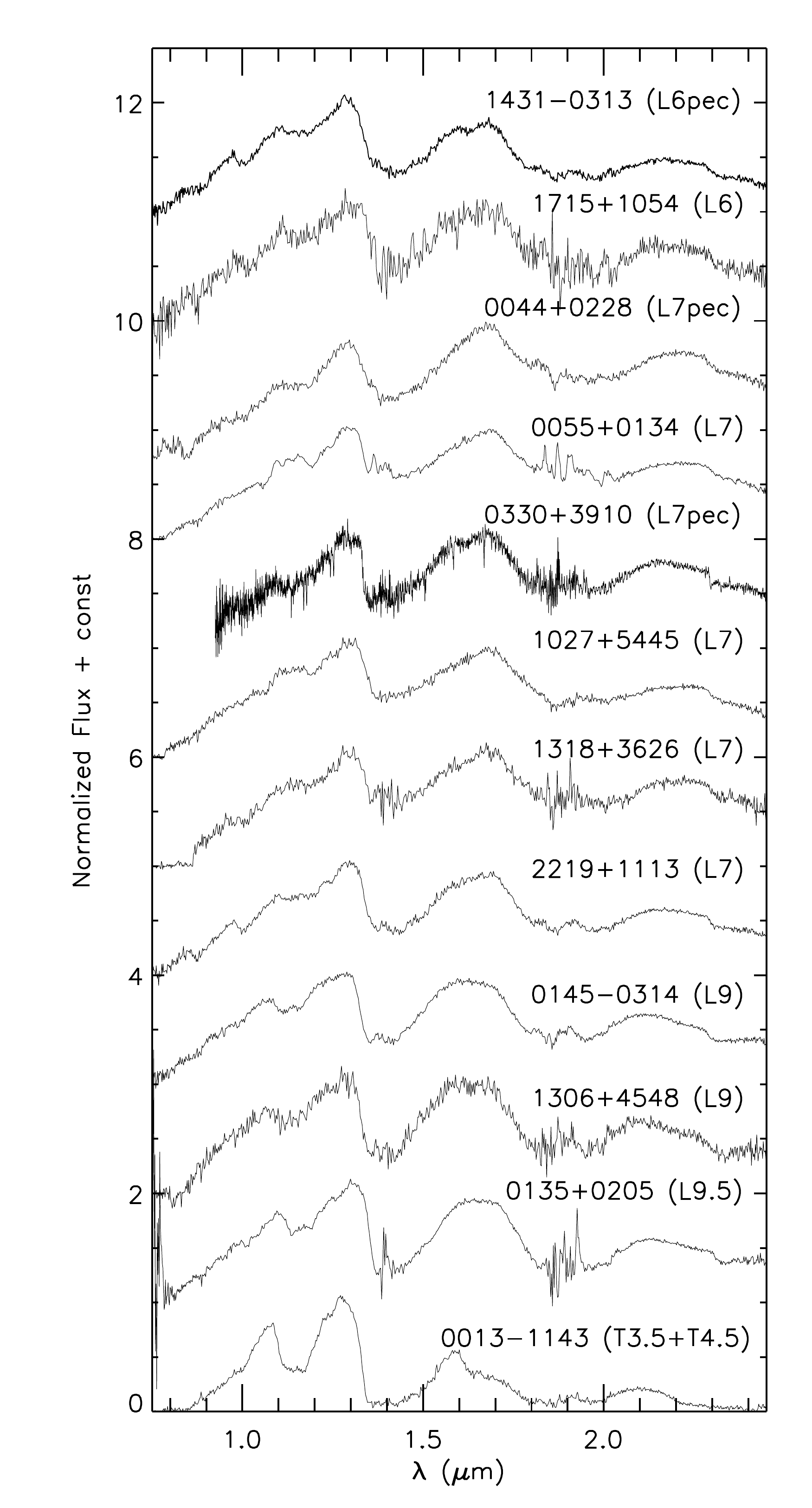}
\end{subfigure}%
\caption{}
\end{figure*}

\renewcommand{\thefigure}{\arabic{figure}}

\section{Spectral Classification Results} \label{sec:specclass}

We estimate spectral types for our candidates by comparing them to ultra-cool dwarf spectral standards in the SpeX Prism Archive\footnote{Standards used for comparison are from \cite{burgasser04,burgasser06,burgasser06a,burgasser07,burgasser07b,burgasser08,cruz04,reid06,chiu06,kirkpatrick10}}. Our spectral type classifications have an average uncertainty of $\pm$1 spectral type. 

\begin{figure*}
\centering
\includegraphics[width=\textwidth]{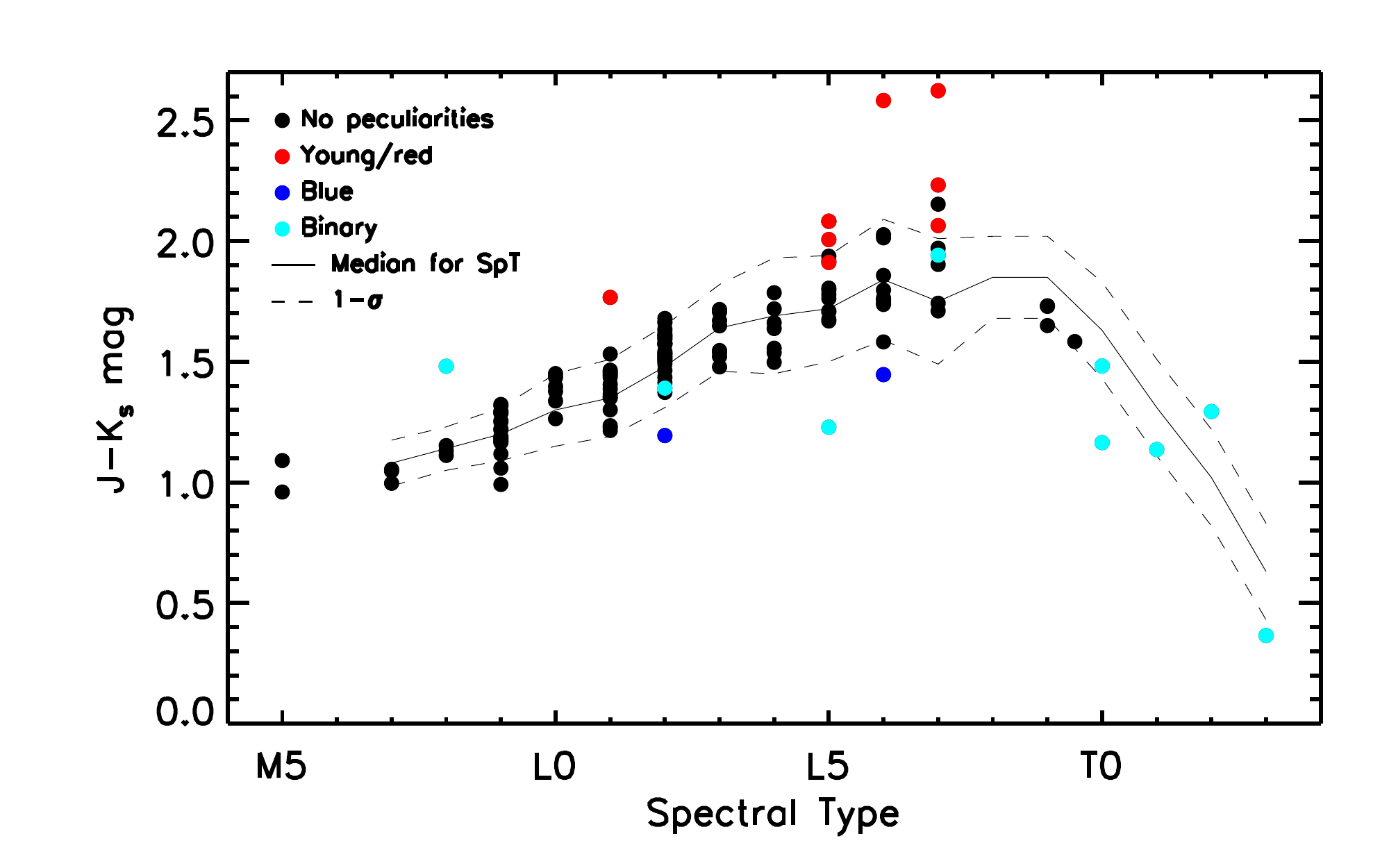}
\caption{Synthetic colors of the 144 new objects discovered in our survey (P1 and present results). The solid line represents the median $J-K_s$ colors for each spectral type from \cite{faherty09} and \cite{faherty13} and the dashed line represents the 1$\sigma$ limits. The correspondence is very good, and indicates that our synthetic colors are reliable for seeking candidate peculiar dwarfs as color outliers.}
\label{fig:sptcol}
\end{figure*}

The newly classified M, L, and T dwarfs are plotted on the $z-J$ vs. $J-K_s$ color-color diagram in Figure \ref{fig:ccdiag}b, where we have used the synthetic colors integrated from the spectra. The spectral types and synthetic colors are presented in Table~\ref{tab:results}. The GNIRS spectra do not cover the entire $z$-band so the $z-J$ colors for the objects taken with GNIRS are their photometric colors. A number of objects have synthetic $z-J$ colors that are bluer than $z-J>$ 2.5~mag. Many of the photometric magnitudes may have been subject to flux-overestimation bias at $J$-band (Section~3.3 of P1) and their colors are close to or below the limits of our $z-J>$ 2.5~mag color selection criteria (Section~\ref{sec:selection}). This is likely the reason why the synthetic and photometric colors are not the same and why some objects no longer satisfy the photometric selection criteria with their synthetic colors (Fig.\ \ref{fig:ccdiag}b). We check that the synthetic colors of our normal L and T dwarfs correctly represent those of the field L and T population by comparing to the L and T near-infrared color compendium of Faherty et al.\ (\citeyear{faherty09,faherty13}).  Figure~\ref{fig:sptcol} shows that there is a very good match.  Therefore, we are confident that our procedure of adopting synthetic colors to correct the low-SNR SDSS and 2MASS photometry can be used to also identify color outliers: candidate peculiar objects.

We discuss the peculiar ($\S$\ref{subsec:pec}), candidate binary ($\S$\ref{subsec:binaries}), and normal ($\S$\ref{subsec:norm}) ultra-cool dwarfs, and false-positives ($\S$\ref{subsec:fp}) in our sample below.

\subsection{Peculiar L Dwarfs} \label{subsec:pec}

We classify seven objects as unusual based on their spectroscopic peculiarities. Our assessment of peculiarity is determined by high spectral similarity to objects that have previously been classified as peculiar. The peculiar characteristics of our objects can be produced by youth, large amounts of atmospheric dust, or low-metallicity. The most interesting objects from this portion of the survey are the young, planetary-mass L7 dwarf 2MASS J00440332+0228112 ($\S$\ref{subsec:0044}), the very red L6 dwarf 2MASS J03530419+0418193 ($\S$\ref{subsec:0353}), and the candidate young L1 dwarf 2MASS J00133470+1109403 ($\S$\ref{subsec:0013}). 

Young ultra-cool dwarfs have low surface gravity, hence, the line strengths of the gravity-sensitive features will differ from those in older objects (e.g. \citealp{lucas01, gorlova03, mcgovern04, allers07, lodieu08, rice10, allers13}). The \ion{Na}{1} (1.138 and 1.141 \micron) and \ion{K}{1} (1.169 and 1.178 \micron, 1.244 and 1.253 \micron) doublets are weaker because of decreased pressure broadening. The FeH features (bandheads at 0.990 \micron\ and 1.194 \micron) are weaker because of decreased opacity of the refractory species. Collision-induced absorption from molecular hydrogen also changes as a function of gravity, with lower collision rates in low-gravity objects imparting a triangular shape to the $H$-band. 

As discussed in P1, the indices developed by \cite{allers13}, \cite{canty13} and \cite{schneider14} have the potential to offer a quantitative gravity classification. However, our spectra have significantly lower spectral resolution, so the index measurements are more uncertain than in those studies.  In addition, most of the indices do not extend into the late-L dwarfs, and so are inadequate to classify some of our most interesting objects.  Therefore, we do not adopt spectral indices as a default gravity classification scheme.  However, we do check for consistency with applicable spectral indices whenever we find peculiarities in the spectra of our L and T candidates.

We note that unusually red objects that lack some signatures of youth can still exhibit some of the spectroscopic characteristics of young objects, in particular weaker FeH bands and a triangular $H$-band continuum (e.g. \citealp{looper08b,kirkpatrick10}). In these objects, such features have been attributed to high atmospheric dust content or to thicker clouds rather than to low-gravity \citep{looper08b,allers13}. In the cases of unusually red objects, we rely on the strength of the gravity-sensitive alkali (\ion{K}{1} and \ion{Na}{1}) lines to distinguish between young objects and field-age ($\gtrsim$0.5 Gyr) objects with unusually high dust content.

For objects with blue near-infrared colors, we seek to determine whether the blue colors may result from low cloud opacity, low-metallicity or unresolved binarity. In order to check the possibility of them being binaries, we consider the CH$_{4}$ in the $K$- and $H$-bands. If there is a higher abundance of CH$_{4}$ in the $H$-band relative to the $K$-band, the object is most likely a binary as the onset of methane absorption in cooler, older objects is apparent in the $K$-band before the $H$-band (e.g \citealp{cruz04,burgasser10,bardalez14}). If we determine that the objects are not binaries, then we can also check whether or not they are metal-poor by analyzing the FeH feature in the $Y$- and $J$-bands. Although an object may be metal-poor, it may have a stronger FeH feature due to the larger relative amounts of hydrogen present and the decreased absorption by oxides such as TiO and VO (e.g. \citealp{kirkpatrick10}). If, however, an object's blue colors come from a low cloud opacity, the overall dust continuum will simply be much weaker, leaving CO and CH$_{4}$ as the dominant opacity sources \citep{kirkpatrick05}.

We discuss individual objects and their defining characteristics in the next sections. 

%\floattable
\begin{deluxetable*}{lccccccccc}
%\rotate
\tabletypesize{\scriptsize}
\tablecolumns{10}
\tablewidth{0pt}
\tablecaption{Results from Spectroscopic Classification and Synthetic Photometry
\label{tab:results}}
\tablehead{
\colhead{2MASS ID} & \colhead{IR} & \colhead{Interpretation} & \colhead{$z-J$} & \colhead{$J-H$} & \colhead{$J-K_s$} & \colhead{$>$1$\sigma$ color} & \colhead{$>$2$\sigma$ color} \\
\colhead{(J2000)} & \colhead{SpT} & \colhead{(from spectrum)} & \colhead{(mag)} & \colhead{(mag)} & \colhead{(mag)} & \colhead{Outlier} & \colhead{Outlier} }
\startdata
{\small Peculiarly Red}  & & & & & & & & & \\
\hline
\object{2MASS J00065552+0236376} & L3 &  & 2.80 $\pm$ 0.05 & 1.01 $\pm$ 0.08 & 1.65 $\pm$ 0.08 & & \\
\object{2MASS J00150673+3006004} & L2 &  & 2.86 $\pm$ 0.05 & 0.98 $\pm$ 0.09 & 1.62 $\pm$ 0.07 & & \\
\object{2MASS J00440332+0228112} & L7pec & red/young & 3.21 $\pm$ 0.10 & 1.33 $\pm$ 0.17 & 2.23 $\pm$ 0.10 & $+$ & \\
\object{2MASS J00501561+1012431} & L2 &  & 2.88 $\pm$ 0.10 & 1.00 $\pm$ 0.12 & 1.66 $\pm$ 0.13 & &  \\
\object{2MASS J00550564+0134365} & L7 &  & 2.98 $\pm$ 0.06 & 1.15 $\pm$ 0.07 & 1.90 $\pm$ 0.07 & &  \\
\object{2MASS J01165802+4333081} & L6 &  & 2.83 $\pm$ 0.06 & 1.12 $\pm$ 0.12 & 1.86 $\pm$ 0.10 & &  \\
\object{2MASS J01183399+1810542} & L2 &  & 2.86 $\pm$ 0.05 & 1.01 $\pm$ 0.09 & 1.68 $\pm$ 0.07 & & \\
\object{2MASS J01341675--0546530} & L2 &  & 2.78 $\pm$ 0.04 & 0.94 $\pm$ 0.07 & 1.51 $\pm$ 0.08 & & \\
\object{2MASS J01392388--1845029} & L6 &  & 2.87 $\pm$ 0.06 & 1.08 $\pm$ 0.08 & 1.74 $\pm$ 0.08 & & \\
\object{2MASS J01394906+3427226} & L1 &  & 2.61 $\pm$ 0.11 & 0.94 $\pm$ 0.12 & 1.53 $\pm$ 0.12 & & \\
\object{2MASS J01453520--0314117} & L9 &  & 2.55 $\pm$ 0.10 & 1.09 $\pm$ 0.14 & 1.73 $\pm$ 0.11 & & \\
\object{2MASS J02022917+2305141} & L6 &  & $\cdots$\tablenotemark{a} & 1.18 $\pm$ 0.13 & 2.01 $\pm$ 0.15 &  & \\
\object{2MASS J02151451+0453179} & L2 &  & 2.77 $\pm$ 0.07 & 1.02 $\pm$ 0.12 & 1.66 $\pm$ 0.12 & & \\
\object{2MASS J03530419+0418193} & L6pec & very red & 3.61 $\pm$ 0.06 & 1.59 $\pm$ 0.10 & 2.58 $\pm$ 0.11 & & $+$ \\
\object{2MASS J04214620--0025072} & L2 &  & 2.65 $\pm$ 0.06 & 0.92 $\pm$ 0.09 & 1.51 $\pm$ 0.08 & & \\
\object{2MASS J04234652+0843211} & L2 &  & 2.69 $\pm$ 0.04 & 0.95 $\pm$ 0.09 & 1.46 $\pm$ 0.05 &  & \\
\object{2MASS J07244848+2506143} & L4 &  & 2.82 $\pm$ 0.05 & 0.96 $\pm$ 0.11 & 1.54 $\pm$ 0.10 & & \\
\object{2MASS J07552723+1138485} & L2 &  & 2.91 $\pm$ 0.13 & 0.97 $\pm$ 0.23 & 1.64 $\pm$ 0.15 & & \\
\object{2MASS J08443811+2226161} & L2 &  & 2.78 $\pm$ 0.07 & 0.99 $\pm$ 0.21 & 1.61 $\pm$ 0.13 & & \\
\object{2MASS J09053247+1339138} & L1 &  & 2.64 $\pm$ 0.12 & 0.86 $\pm$ 0.24 & 1.41 $\pm$ 0.23 & & \\
\object{2MASS J09325053+1836485} & L6 &  & 2.80 $\pm$ 0.08 & 1.05 $\pm$ 0.13 & 1.76 $\pm$ 0.20 & & \\
\object{2MASS J09393078+0653098} & L3 &  & 2.74 $\pm$ 0.10 & 1.02 $\pm$ 0.10 & 1.67 $\pm$ 0.11 & & \\
\object{2MASS J10271549+5445175} & L7 &  & 2.86 $\pm$ 0.04 & 1.05 $\pm$ 0.06 & 1.74 $\pm$ 0.06 &  & \\
\object{2MASS J10524963+1858151} & L5pec & red & $\cdots$\tablenotemark{a} & 1.22 $\pm$ 0.23 & 2.08 $\pm$ 0.15 & $+$ &  \\
\object{2MASS J11220855+0343193} & L5 &  & 3.10 $\pm$ 0.05 & 1.04 $\pm$ 0.06 & 1.71 $\pm$ 0.06 & & \\
\object{2MASS J11285958+5110202} & L6 &  & 2.85 $\pm$ 0.03 & 0.99 $\pm$ 0.08 & 1.58 $\pm$ 0.07 & & \\
\object{2MASS J12023885+5345384} & L2 &  & 2.80 $\pm$ 0.12 & 0.98 $\pm$ 0.21 & 1.60 $\pm$ 0.16 & & \\
\object{2MASS J12352675+4124310} & L5 &  & 2.62 $\pm$ 0.08 & 1.01 $\pm$ 0.08 & 1.65 $\pm$ 0.08 & & \\
\object{2MASS J13042886--0032410} & L6 &  & 2.80 $\pm$ 0.08 & 1.08 $\pm$ 0.11 & 1.80 $\pm$ 0.15 & & \\
\object{2MASS J13184567+3626138} & L7 &  & 3.18 $\pm$ 0.09 & 1.18 $\pm$ 0.13 & 1.97 $\pm$ 0.12 &  & \\
\object{2MASS J13451417+4757231} & L3 & & 2.88 $\pm$ 0.06 & 1.05 $\pm$ 0.11 & 1.71 $\pm$ 0.10 & & \\
\object{2MASS J14554511+3843329} & L2 &  & 2.55 $\pm$ 0.06 & 0.90 $\pm$ 0.15 & 1.52 $\pm$ 0.16 & & \\
\object{2MASS J15102256--1147125} & L2 &  & 2.68 $\pm$ 0.04 & 0.88 $\pm$ 0.05 & 1.44 $\pm$ 0.07 & & \\
\object{2MASS J15163838+3333576} & L1 &  & 2.63 $\pm$ 0.09 & 0.89 $\pm$ 0.12 & 1.46 $\pm$ 0.09 & & \\
\object{2MASS J15442544+0750572} & L2 &  & 2.59 $\pm$ 0.09 & 0.98 $\pm$ 0.10 & 1.57 $\pm$ 0.15 & & \\
\object{2MASS J15500191+4500451} & L5 &  & 2.92 $\pm$ 0.10 & 1.10 $\pm$ 0.13 & 1.80 $\pm$ 0.10 &  & \\
\object{2MASS J15543602+2724487} & L5 &  & 2.91 $\pm$ 0.04 & 1.05 $\pm$ 0.09 & 1.71 $\pm$ 0.07 & & \\
\object{2MASS J15565004+1449081} & L5 &  & 2.87 $\pm$ 0.10 & 1.02 $\pm$ 0.16 & 1.67 $\pm$ 0.15 & &  \\
\object{2MASS J17120142+3108217} & L3 &  & 2.84 $\pm$ 0.04 & 1.04 $\pm$ 0.07 & 1.72 $\pm$ 0.07 & & \\
\object{2MASS J17153111+1054108} & L6 & & 2.80 $\pm$ 0.11 & 1.09 $\pm$ 0.15 & 1.76 $\pm$ 0.13 &  & \\
\object{2MASS J17164469+2302220} & L2 &  & $\cdots$\tablenotemark{a} & 0.88 $\pm$ 0.15 & 1.57 $\pm$ 0.15 & &   \\
\object{2MASS J17440969+5135032} & L1 & & 2.62 $\pm$ 0.07 & 0.80 $\pm$ 0.14 & 1.44 $\pm$ 0.12 &  & \\
\object{2MASS J22035781+0713492} & L1 &  & 2.80 $\pm$ 0.05 & 0.72 $\pm$ 0.12 & 1.23 $\pm$ 0.11 & & \\
\object{2MASS J22191282+1113405} & L7 &  & 2.84 $\pm$ 0.08 & 1.05 $\pm$ 0.10 & 1.71 $\pm$ 0.10 & & \\
\object{2MASS J22295358+1556180} & L2 &  & 2.82 $\pm$ 0.06 & 0.96 $\pm$ 0.09 & 1.60 $\pm$ 0.10 & & \\
\object{2MASS J22355244+0418563} & L3 &  & 2.76 $\pm$ 0.03 & 0.92 $\pm$ 0.03 & 1.48 $\pm$ 0.04 & & \\
\object{2MASS J23313131+2041273} & L4 &  & 2.78 $\pm$ 0.05 & 0.97 $\pm$ 0.08 & 1.56 $\pm$ 0.09 & & \\
\hline
{\small Candidate Binary} & & & & & & & & & \\
\hline
\object{2MASS J00100480--0930519} & M9 &  & 2.40 $\pm$ 0.05 & 0.67 $\pm$ 0.22 & 1.12 $\pm$ 0.16 & & \\
\object{2MASS J01114355+2820024} & M9 &  & 2.38 $\pm$ 0.06 & 0.72 $\pm$ 0.17 & 1.22 $\pm$ 0.18 & & \\
\object{2MASS J01141304+4354287} & M5 &  & 1.74 $\pm$ 0.09 & 0.65 $\pm$ 0.17 & 0.96 $\pm$ 0.19 & & \\
\object{2MASS J01414428+2227409} & L3 &  & 2.72 $\pm$ 0.11 & 0.95 $\pm$ 0.13 & 1.55 $\pm$ 0.24 & & \\
\object{2MASS J09194512+5135149} & M9 &  & 2.40 $\pm$ 0.05 & 0.71 $\pm$ 0.13 & 1.18 $\pm$ 0.16 & & \\
\object{2MASS J10592523+5659596} & L1 &  & 2.66 $\pm$ 0.05 & 0.84 $\pm$ 0.15 & 1.35 $\pm$ 0.12 & & \\
\object{2MASS J12453705+4028456} & L1 &  & 2.59 $\pm$ 0.04 & 0.72 $\pm$ 0.17 & 1.22 $\pm$ 0.18 & & \\
\object{2MASS J13170488+3447513} & L0 &  & 2.53 $\pm$ 0.04 & 0.80 $\pm$ 0.15 & 1.26 $\pm$ 0.18 & & \\
\object{2MASS J14124574+3403074} & L0 &  & 2.54 $\pm$ 0.06 & 0.80 $\pm$ 0.12 & 1.34 $\pm$ 0.15 & & \\
\object{2MASS J14313545--0313117} & L6pec & blue & 2.78 $\pm$ 0.04 & 0.88 $\pm$ 0.10 & 1.45 $\pm$ 0.12 & $-$ & \\
\object{2MASS J15525579+1123523} & M9 &  & 2.48 $\pm$ 0.04 & 0.79 $\pm$ 0.10 & 1.29 $\pm$ 0.13 & & \\
\object{2MASS J21123034+0758505} & M9 &  & 2.53 $\pm$ 0.04 & 0.78 $\pm$ 0.16 & 1.29 $\pm$ 0.16 & & \\
\object{2MASS J22582325+2906484} & M8 &  & 2.22 $\pm$ 0.05 & 0.67 $\pm$ 0.19 & 1.11 $\pm$ 0.22 & & \\
\hline
{\small Peculiarly Red and Candidate Binary}  & & & & & & & & & \\
\hline
\object{2MASS J00082822+3125581} & M7 &  & 2.18 $\pm$ 0.14 & 0.63 $\pm$ 0.08 & 1.00 $\pm$ 0.08 & & \\
\object{2MASS J00132229--1143006} & T3pec & T3.5 + T4.5? & 3.31 $\pm$ 0.10 & 0.34 $\pm$ 0.11 & 0.37 $\pm$ 0.22 & $-$ & \\
\object{2MASS J00133470+1109403} & L1pec & red/young & 2.94 $\pm$ 0.04 & 1.06 $\pm$ 0.06 & 1.77 $\pm$ 0.07 & $+$ & \\
\object{2MASS J00452972+4237438} & M8pec & M8 + L7? & 2.71 $\pm$ 0.13 & 0.96 $\pm$ 0.17 & 1.48 $\pm$ 0.14 & & \\
\object{2MASS J01145788+4318561} & M9 &  & 2.44 $\pm$ 0.11 & 0.64 $\pm$ 0.03 & 1.06 $\pm$ 0.04 & & \\
\object{2MASS J01194279+1122427} & M8 &  & 2.23 $\pm$ 0.08 & 0.69 $\pm$ 0.10 & 1.13 $\pm$ 0.12 & & \\
\object{2MASS J01412651+1001339} & L5 &  & $\cdots$\tablenotemark{a} & 0.93 $\pm$ 0.17 & 1.92 $\pm$ 0.21 & & \\
\object{2MASS J01442482--0430031} & L4 & & 2.86 $\pm$ 0.16 & 1.07 $\pm$ 0.20 & 1.79 $\pm$ 0.19 & & \\
\object{2MASS J02314893+4521059} & M9 &  & 2.46 $\pm$ 0.14 & 0.74 $\pm$ 0.12 & 0.99 $\pm$ 0.10 & $-$ & \\
\object{2MASS J03302948+3910242} & L7pec & red & $\cdots$\tablenotemark{a} & 1.25 $\pm$ 0.20 & 2.06 $\pm$ 0.16 & $+$ & \\
\object{2MASS J03315828+4130486} & M5 &  & 1.79 $\pm$ 0.15 & 0.74 $\pm$ 0.21 & 1.09 $\pm$ 0.26 & & \\
\object{2MASS J09240328+3653444} & L2pec & L2 + T3? & $\cdots$\tablenotemark{a} & 0.79 $\pm$ 0.20 & 1.39 $\pm$ 0.24 & $-$ &  \\
\object{2MASS J10265851+2515262} & L5pec & red & $\cdots$\tablenotemark{a} & 1.15 $\pm$ 0.20 & 2.01 $\pm$ 0.19 & $+$ &  \\
\object{2MASS J11060459--1907025} & L5 &  & 2.93 $\pm$ 0.12 & 1.10 $\pm$ 0.17 & 1.78 $\pm$ 0.13 & &  \\
\object{2MASS J12232570+0448277} & L6 &  & 2.86 $\pm$ 0.05 & 1.09 $\pm$ 0.11 & 1.74 $\pm$ 0.13 & & \\
\object{2MASS J13064517+4548552} & L9 &  & 2.90 $\pm$ 0.14 & 1.14 $\pm$ 0.20 & 1.73 $\pm$ 0.16 & & \\
\object{2MASS J14193789+3333326} & L3 &  & $\cdots$\tablenotemark{a} & 1.07 $\pm$ 0.10 & 1.52 $\pm$ 0.10 & & \\
\object{2MASS J17570962+4325139} & L2 &  & 2.77 $\pm$ 0.08 & 0.93 $\pm$ 0.21 & 1.57 $\pm$ 0.13 & & \\
\object{2MASS J23004298+0200145} & L2 & & 2.89 $\pm$ 0.05 & 0.98 $\pm$ 0.11 & 1.63 $\pm$ 0.13 & &  \\
\object{2MASS J23053808+0524070} & L2 &  & 2.74 $\pm$ 0.08 & 0.95 $\pm$ 0.14 & 1.58 $\pm$ 0.10 & & \\
\hline
{\small General Ultra-cool Dwarf Candidates}  & & & & & & & & & \\
\hline
\object{2MASS J00062250+1300451} & L2 &  & 2.71 $\pm$ 0.06 & 0.95 $\pm$ 0.12 & 1.53 $\pm$ 0.18 & & \\
\object{2MASS J00435012+0928429} & L2 &  & 2.83 $\pm$ 0.03 & 1.00 $\pm$ 0.15 & 1.59 $\pm$ 0.12 & & \\
\object{2MASS J01001471--0301494} & L1 &  & 2.64 $\pm$ 0.05 & 0.85 $\pm$ 0.12 & 1.45 $\pm$ 0.12 & & \\
\object{2MASS J01343635--0145444} & L0 &  & 2.56 $\pm$ 0.05 & 0.86 $\pm$ 0.10 & 1.40 $\pm$ 0.14 & & \\
\object{2MASS J01352531+0205232} & L9.5 &  & 2.81 $\pm$ 0.05 & 1.01 $\pm$ 0.10 & 1.58 $\pm$ 0.12 &  & \\
\object{2MASS J03511847--1149326} & L2 &  & 2.68 $\pm$ 0.05 & 0.92 $\pm$ 0.09 & 1.54 $\pm$ 0.11 & & \\
\object{2MASS J04234652--0803051} & L0 & & 2.33 $\pm$ 0.04 & 0.85 $\pm$ 0.09 & 1.45 $\pm$ 0.13 & & \\
\object{2MASS J04510592+0014394} & M9 & & 2.57 $\pm$ 0.08 & 0.74 $\pm$ 0.17 & 1.25 $\pm$ 0.17 & & \\
\object{2MASS J08270185+4129191} & L1 &  & 2.67 $\pm$ 0.03 & 0.86 $\pm$ 0.10 & 1.39 $\pm$ 0.08 & & \\
\object{2MASS J09083688+5526401} & L1 &  & 2.71 $\pm$ 0.05 & 0.88 $\pm$ 0.13 & 1.46 $\pm$ 0.09 & & \\
\object{2MASS J09481259+5300387} & L2 &  & 2.71 $\pm$ 0.04 & 0.85 $\pm$ 0.07 & 1.37 $\pm$ 0.06 & & \\
\object{2MASS J10551343+2504028} & L2 &  & 2.68 $\pm$ 0.08 & 0.96 $\pm$ 0.11 & 1.61 $\pm$ 0.15 & & \\
\object{2MASS J11213919--1053269} & L1 &  & 2.77 $\pm$ 0.04 & 0.86 $\pm$ 0.09 & 1.44 $\pm$ 0.14 & & \\
\object{2MASS J11282763+5934003} & L0 &  & 2.48 $\pm$ 0.05 & 0.84 $\pm$ 0.16 & 1.38 $\pm$ 0.13 & & \\
\object{2MASS J12260640+1756293} & L2 &  & $\cdots$\tablenotemark{a} & 0.80 $\pm$ 0.19 & 1.41 $\pm$ 0.21 &  &  \\
\object{2MASS J12492272+0310255} & L2 &  & 2.74 $\pm$ 0.04 & 0.93 $\pm$ 0.12 & 1.54 $\pm$ 0.13 & & \\
\object{2MASS J13264464+3627407} & L2 &  & 2.56 $\pm$ 0.06 & 0.93 $\pm$ 0.10 & 1.52 $\pm$ 0.10 & & \\
\object{2MASS J14154242+2635040} & L0 &  & 2.68 $\pm$ 0.05 & 0.85 $\pm$ 0.15 & 1.43 $\pm$ 0.11 & & \\
\object{2MASS J15025475+5044252} & L2 & & $\cdots$\tablenotemark{a} & 0.97 $\pm$ 0.13 & 1.50 $\pm$ 0.15 & & \\
\object{2MASS J15202471+2203340} & L1 & & $\cdots$\tablenotemark{a} & 0.78 $\pm$ 0.17 & 1.36 $\pm$ 0.20 & & \\
\object{2MASS J15552840+5918155} & L4 & & $\cdots$\tablenotemark{a} & 1.05 $\pm$ 0.10 & 1.72 $\pm$ 0.12 & & \\
\object{2MASS J16123860+3126489} & L2 &  & 2.64 $\pm$ 0.05 & 0.94 $\pm$ 0.16 & 1.54 $\pm$ 0.13 & & \\
\object{2MASS J16194822--0425366} & L1 & & $\cdots$\tablenotemark{a} & 0.76 $\pm$ 0.14 & 1.23 $\pm$ 0.18 & & \\
\object{2MASS J22545900--0330590} & M9 &  & 2.41 $\pm$ 0.08 & 0.70 $\pm$ 0.16 & 1.17 $\pm$ 0.17 & & \\
\hline
{\small False-positives}  & & & & & & & & & \\
\hline
\object{2MASS J01581172+3232013} & ? & & 1.03 $\pm$ 0.04 & 0.50 $\pm$ 0.07 & 0.59 $\pm$ 0.05 \\
\object{2MASS J02555058+1926476} & ? & & $\cdots$\tablenotemark{a} & 2.62 $\pm$ 0.10 & 3.69 $\pm$ 0.10 \\
\object{2MASS J02553101+1929356} & ? & & $\cdots$\tablenotemark{a} & 1.66 $\pm$ 0.13 & 2.46 $\pm$ 0.12 \\
\object{2MASS J04084337+5120524} & ? & & 3.49 $\pm$ 0.06 & 1.87 $\pm$ 0.04 & 3.13 $\pm$ 0.04 \\
\object{2MASS J05484895+0014367} & ? & & 3.62 $\pm$ 0.06 & 1.95 $\pm$ 0.03 & 2.79 $\pm$ 0.03 \\
\object{2MASS J05480405+0029264} & ? & & 3.27 $\pm$ 0.10 & 1.75 $\pm$ 0.06 & 2.48 $\pm$ 0.08 \\
\object{2MASS J05584262+2150121} & ? & & $\cdots$\tablenotemark{b} & 2.01 $\pm$ 0.04 & 2.97 $\pm$ 0.02 \\
\object{2MASS J06380876+0940084} & ? & & 3.20 $\pm$ 0.16 & 1.95 $\pm$ 0.06 & 3.30 $\pm$ 0.04 \\
\object{2MASS J06415196+0916111} & ? & & 3.65 $\pm$ 0.08 & 1.87 $\pm$ 0.07 & 2.71 $\pm$ 0.05 \\
\object{2MASS J09232257+5208598} & ? & & 0.97 $\pm$ 0.05 & 0.43 $\pm$ 0.28 & 0.49 $\pm$ 0.20 \\
\object{2MASS J16472470--0935294} & ? & & 4.34 $\pm$ 0.14 & 2.34 $\pm$ 0.12 & 3.48 $\pm$ 0.10 \\
\object{2MASS J16484099+2231397} & ? & & 4.34 $\pm$ 0.05 & 2.35 $\pm$ 0.08 & 3.48 $\pm$ 0.09 \\
\enddata
\tablecomments{We identify color outliers by comparing the synthetic $J-K_s$ color of each object to the median $J-K_s$ colors of M7--M9 and T0--T8 dwarfs from \cite{faherty09} and for L0--L9 dwarfs from \cite{faherty13}. The $+$ and $-$ signs indicate whether the object is above or below the average, respectively. The objects were divided into the different categories based on their photometric colors. The objects in the ``Peculiarly Red and Candidate Binary" category passed both the peculiarly red and candidate L+T binary selection criteria.}
\tablenotetext{a}{\footnotesize The spectra of these objects are from GNIRS, and do not cover the entire SDSS $z$-band, so we are unable to calculate synthetic $z-J$ colors.}
\tablenotetext{b}{\footnotesize The spectra of these objects are from Magellan/FIRE. The FIRE prism spectra do not cover the entire SDSS $z$-band so we are unable to calculate the $z-J$ colors.}
\end{deluxetable*}

\subsubsection{2MASS J00133470+1109403 (L1)} \label{subsec:0013}
This object is a young L1 dwarf. Compared to a normal L1.5 dwarf, 2M J0013+1109 has much weaker FeH and \ion{K}{1} absorption lines and a triangular $H$-band (Fig.\ \ref{fig:pec1}). The \cite{allers13} spectral indices say that this object is an INT-G object. This object is very similar to the L1 $\beta$ brown dwarf 2MASS J01174748--3403258 \citep{burgasser08}, however, it has even weaker \ion{K}{1} absorption lines and a redder continuum. Based on these characteristics, 2M 0013+1109 likely has lower gravity than 2M 0117--3403 and so could potentially be a $\sim$10 Myr-old free-floating planetary-mass object.

However, according to BANYAN II \citep{malo13,gagne14} and the Convergent Point tool \citep{rodriguez13}, 2M 0013+1109 does not have a likelihood of being a part of any of the young associations and moving groups used in those works. Instead, BANYAN II calculates a 98\% probability of being part of the young field when using a young prior, and 60\% probability of being part of the young field when no priors are set.

\subsubsection{2MASS J00440332+0228112 (L7)} \label{subsec:0044}
This object is a young L7 brown dwarf based on weak \ion{K}{1} and FeH absorption, stronger $J$-band absorption of H$_2$O and a more triangular shaped $H$-band compared to the normal L7 dwarf 2MASS J0028208+224905 (\citealp{burgasser10}; Fig.\ \ref{fig:pec1}). The gravity-sensitive features in the $J$-band and the shape of the $H$-band are similar to those in the young L7 dwarf 2MASSI J0103320+193536 \citep{cruz04}, although the observed spectrum is slightly redder than the comparison spectrum in Fig.\ \ref{fig:pec1}. The only spectral index available for such a late spectra type from \cite{allers13} and \cite{schneider14} is the $H$-cont index which indicates a VL-G brown dwarf.  However, as noted by \cite{allers13}, very red L dwarfs with no youth signatures can also exhibit triangular $H$-band shapes. In this case we can only say that the spectral index is consistent with the result from the spectral comparison.

This object was also independently reported as a young L7 dwarf in \cite{schneider17}. They determine that there is a high probability that it belongs to the $\beta$ Pictorus Moving Group according to BANYAN II (\citealp{malo13,gagne14}; $\sim$78\%) and the Convergent Point tool (\citealp{rodriguez13}; $\sim$97\%). They also report that based on its photometric distance (31 $\pm$ 3 pc), age (24 $\pm$ 3 Myr), and bolometric luminosity, it has a mass range of 7--11 M$_{Jup}$.

\subsubsection{2MASS J03302948+3910242 (L7)}
This object is a peculiarly red L7 dwarf. The spectrum of 2M 0330+3910 is similar to both L6 and L7 dwarfs in the $J$- and $H$-bands but the shape of the $K$-band is more similar to an L7 dwarf. The gravity-sensitive \ion{K}{1} features are not weaker than that of a normal L7 object and the $H$-band does not have the characteristic triangular shape (Fig.\ \ref{fig:pec1}) so this object is not young. This object more closely resembles the peculiarly red L6 brown dwarf 2MASS J21481633+4003594 \citep{kirkpatrick10} even though 2M J0330+3910 is not as red. The $J$-band absorption features also do not quite match but that could be attributed to the difference in spectral type and the slope of the continuum. The spectral indices are also consistent with this object being a FLD-G object.

\subsubsection{2MASS J03530419+0418193 (L6)} \label{subsec:0353}
Not only is this one of the reddest objects observed to date ($z-J=$ 3.61 $\pm$ 0.06~mag; $J-K_s =$ 2.58 $\pm$ 0.11~mag), it is also one of the reddest known objects that does not have any signatures of youth. The strength of the gravity-sensitive \ion{K}{1} absorption lines are comparable to those of a peculiarly red L6 dwarf (Fig.\ \ref{fig:pec1}). The $H$-band continuum also does not have the characteristic sharp triangular shape of a young object. The strength of the 0.99$\micron$ FeH feature appears to be decreased, however, this could be a result of the extremely red continuum slope as it is more comparable to the FeH strength of the peculiarly red L6 object 2MASS J21481633+4003594 \citep{kirkpatrick10}. The strength of the other diagnostic absorption features and the shape of the $H$-band continuum are also consistent with those of the peculiarly red L6 dwarf. As with 2M J0044+0228, the only spectral index available for this object is the $H$-cont index so we cannot use spectral indices to help us determine surface gravity in this case. 

Previously, the reddest observed object that had been confirmed with no signatures of youth was the L7 dwarf WISE J233527.07+451140.9, at $J-K_s =$ 2.54 $\pm$ 0.05~mag \citep{liu16}. This object was discovered by \cite{thompson13} in a search for ultra-cool dwarf members of the solar neighborhood using photometry from 2MASS and the WISE All-Sky Source Catalog. The late-L/early-T dwarf WISE J173859.27+614242.1 also has an extremely red color (2MASS $J-K_s =$ 2.55 $\pm$ 0.16~mag; \citealp{mace13}) which has been speculated to be caused from something other than youth but its spectral type and relative surface gravity are still not known.

\subsubsection{2MASS J10265851+2515262 (L5)}
This object is a peculiarly red L5 dwarf. The GNIRS spectrum in Figure \ref{fig:pec1} has been smoothed to the same resolution as the SpeX spectra for more direct comparison. We note that the GNIRS spectrum has a low SNR ($\sim$7--15) so definitive determination of the absorption strengths cannot be attained. We can see, however, that the \ion{K}{1} absorption lines appear to have the same or greater strength than those of the normal L5 comparison object and the peculiarly red L5 (2MASS J23512200+3010540; \citealp{kirkpatrick10}) and that the $H$-band doesn't appear to be triangular in shape (Fig.\ \ref{fig:pec1}). Overall, the spectrum of 2M 1026+2515 is more similar to the red L5 dwarf. The \cite{allers13} indices say that this object is a FLD-G object but the low SNR makes this determination unreliable. However, this object is clearly red so together with the lack of reduced absorption strength of the gravity-sensitive features, the spectroscopic features point to this object being a peculiarly red L5 dwarf.

\subsubsection{2MASS J10524963+1858151 (L5)}
This object is also a peculiarly red L5 dwarf. The GNIRS spectrum in Figure \ref{fig:pec1} has been smoothed to the same resolution as the SpeX spectra for more direct comparison. The spectral indices classify this object as an INT-G object most likely because of its very red color, however, the comparison to other L5 objects is inconsistent with this classification. The \ion{K}{1} and FeH absorption features are similar in strength to those of a normal L5 object and the peculiarly red L5 2MASS J23512200+3010540 \citep{kirkpatrick10}. The $H$-band continuum also does not resemble that of a young object. We note that the GNIRS spectrum has relatively low signal-to-noise (SNR$\sim$10--20) which makes the index values highly uncertain.

\begin{figure*}
\centering
\includegraphics[scale=0.65]{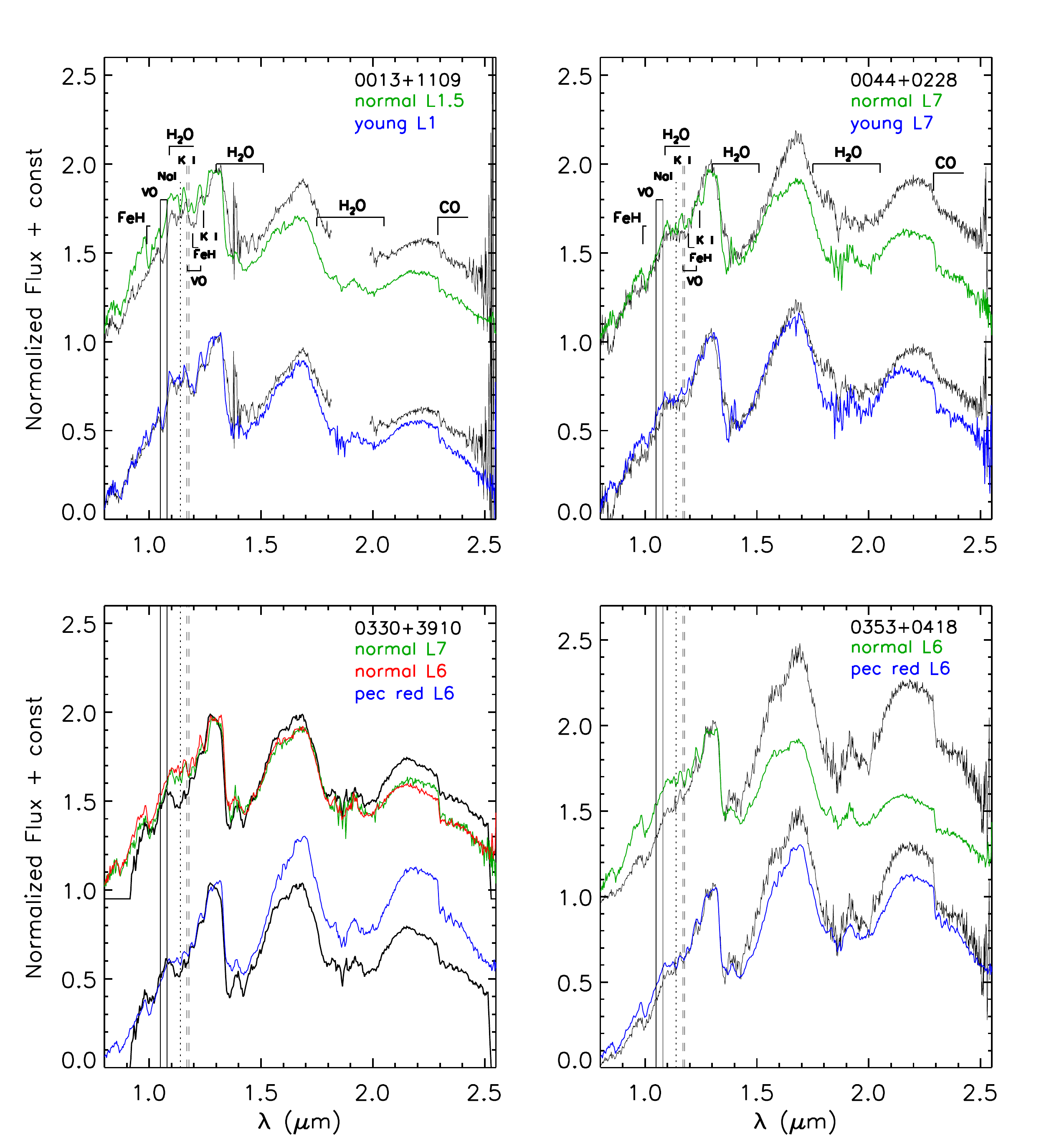}
\caption{Spectra of the peculiar single objects identified in this work. The spectra are compared to spectra of both a normal object of the same spectral type and of a peculiar object of the nearest spectral type. The comparison spectra from left to right and top to bottom are: L1.5 (2MASS J07415784+0531568; \citealp{kirkpatrick10}) and young L1 (2MASSI J0117474$-$340325; \citealp{burgasser08}); L7 (2MASS J0028208+224905; \citealp{burgasser10}) and young L7 (2MASSI J0103320+193536; \citealp{cruz04}); L7 (2MASS J0028208+224905; \citealp{burgasser10}), L6 (2MASSI J1010148$-$040649; \citealp{reid06}) and red L6 (2MASS J21481633+4003594; \citealp{kirkpatrick10}); L6 (2MASSI J1010148$-$040649; \citealp{reid06}) and red L6 (2MASS J21481633+4003594; \citealp{kirkpatrick10}).}
\label{fig:pec1}
\end{figure*}

\subsubsection{2MASS J14313545$-$0313117 (L6)}
This object is an unusually blue L6 dwarf. The spectrum is much bluer than a normal L6 dwarf and the continuum slope is more similar to the blue L6 dwarf 2MASS J11181292$-$0856106 (\citealp{kirkpatrick10}; Fig.\ \ref{fig:pec1}). However, the absorption features do not match 2M J1118$-$0856. In particular, the \ion{K}{1} absorption features are extremely weak in 2M J1431$-$0313 and the 2.3$\micron$ CH$_4$ feature is much less sharp. The FeH absorption is not noticeably stronger than that of a normal object as might be expected in a low-metallicity object. \cite{kirkpatrick10} discuss a category of objects in which 2M J1118$-$0856 may fall, that are potentially slightly metal-poor but not so much that they are categorized as subdwarfs. These objects do not show signs of significantly reduced metal content but have a higher transverse velocity than the field L dwarf population indicating that they are slightly older. We do not know the transverse velocity of 2M J1431$-$0313 but since its spectral characteristics are similar to those of 2M J1118$-$0856, we tentatively adopt the slightly metal-poor classification for this object as well.

\renewcommand{\thefigure}{\arabic{figure} (cont.)}

\begin{figure*}
\setcounter{figure}{2}
\centering
\includegraphics[scale=0.65]{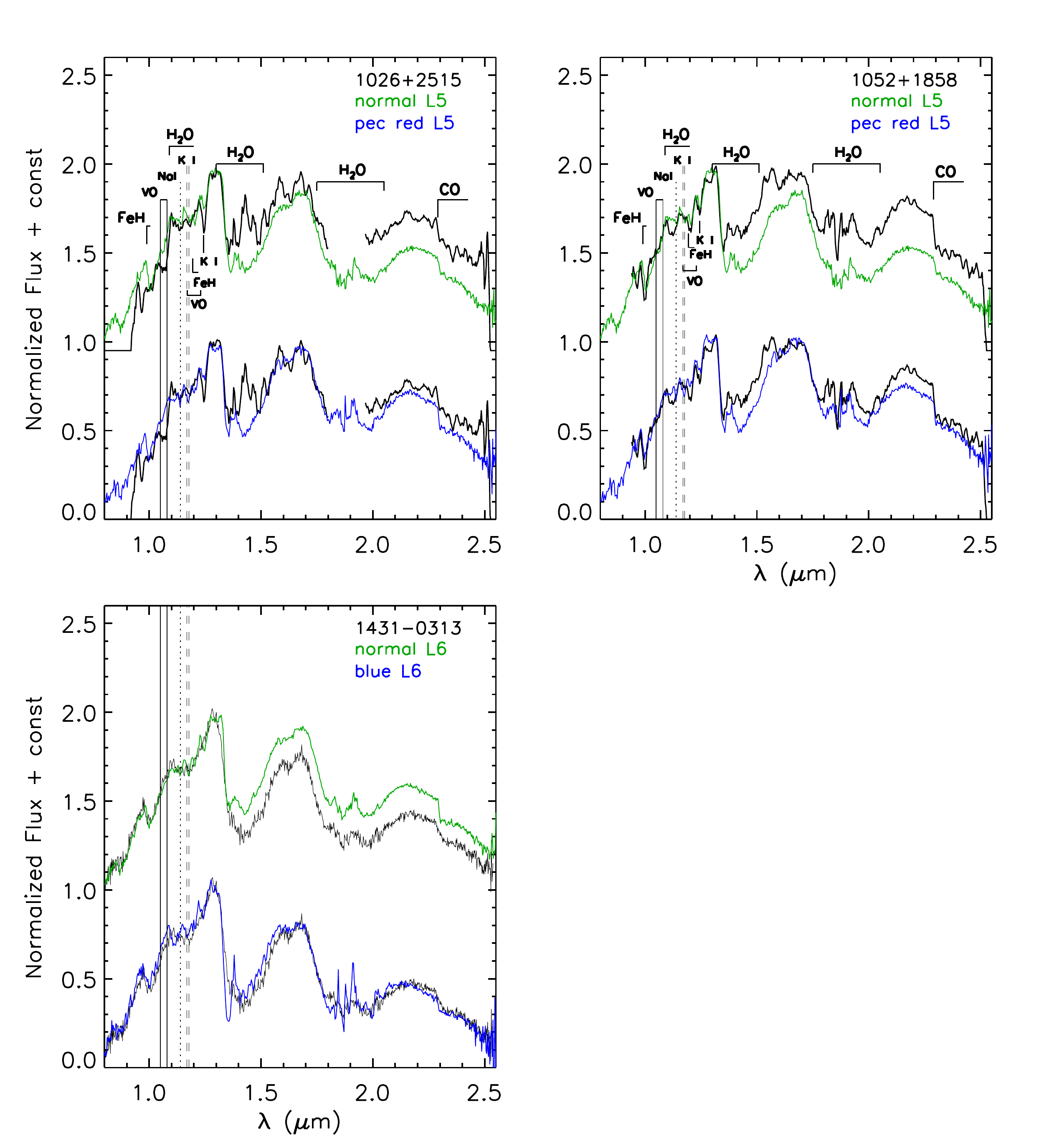}
\caption{The comparison spectra from left to right and top to bottom are: L5 (2MASS J01550354+0950003; \citealp{burgasser10}) and red L5 (2MASS J23512200+3010540; \citealp{kirkpatrick10}); L5 (2MASS J01550354+0950003; \citealp{burgasser10}) and red L5 (2MASS J23512200+3010540; \citealp{kirkpatrick10}); L6 (2MASSI J1010148$-$040640; \citealp{reid06}) and blue L6 (2MASS J11181292$-$0856106; \citealp{kirkpatrick10}).}
\label{fig:pec2}
\end{figure*}

\renewcommand{\thefigure}{\arabic{figure}}

\subsection{Brown Dwarfs with Composite Spectral Types} \label{subsec:binaries}

Three of the objects show peculiarities that do not readily match those found in other individual objects.  Instead, they more closely resemble combination spectra of L and T dwarfs. \cite{burgasser07} and \cite{burgasser10} developed a technique that enables one to infer the spectral types of the individual components of a candidate unresolved binary by a goodness-of-fit comparison to a library of spectral template combinations.  We adopt this technique by creating combination templates from the set of single L and T dwarfs from the SpeX Prism Library. In P1 we created templates by using only the near-infrared standard objects. We have now expanded our templates to include objects from the entire SpeX Prism Library as done in \cite{bardalez14}. 

We constructed our composite template spectra in the same way as P1 but we have now used the updated absolute spectral-type dependent magnitude polynomials given in Table 10 of \cite{filippazzo15}. We classified an object as a likely spectral type composite --- a potential unresolved binary --- if the $\chi^2$ (calculated over the entire 0.8--2.5 \micron\ region, minus the water absorption bands) of the dual-template spectral fit is significantly lower than the $\chi^2$ of the single-template fit.

In addition to template fitting, where applicable, we analyzed the spectral indices defined specifically for SpeX prism spectra in Burgasser et al.\ (\citeyear{burgasser10}; for L+T binaries) and Bardalez Gagliuffi et al.\ (\citeyear{bardalez14}; for M+T and L+T binaries).

\subsubsection{2MASS J00132229--1143006 (T3.5+T4.5?)}
While this object has a $J$-band spectrum consistent with that of a T3 dwarf (Fig.\ \ref{fig:bin1}), it is better fit by a composite template of a T3.5 + T4.5 dwarf. However, this object could instead be a blue T3 dwarf as all the features are similar to those of the T3 dwarf except the blue color. A third explanation is that this object is a variable T dwarf that displays two distinct temperature components. An example is 2MASS J21392676+0220226 which was originally thought to be a candidate L8.5+T3.5 unresolved binary \citep{burgasser10} but was later identified as a high-amplitude variable \citep{radigan12}. This may also be the case with 2M J0013$-$1143. This object only satisfies two of the six binary index selection criteria from Table 5 of \cite{burgasser10}. Overall because of the much better binary template fit, we treat 2M J0013--1143 as a likely unresolved binary but note that it can instead be a blue T3 dwarf.

\subsubsection{2MASS J00452972+4237438 (M8+L7?)}
This object's spectrum is quite unusual and is best fit by a composite template of an M8 and an L7 dwarf (Fig.\ \ref{fig:bin1}). The overall spectral slope is quite red but there are none of the typical spectral features found in a normal ultra-cool dwarf. The spectrum appears to be more similar to a late M dwarf apart from the red slope. However, it does not match any of the late M comparison spectra. The composite template in Figure \ref{fig:bin1} more closely matches the spectrum of 2M J0045+4237 but still does not reproduce all of the features. We note that in the raw data, the profile of the spectra appear to be double-peaked, indicating that this object may be resolvable at higher angular resolution or that there is a contaminating foreground or background object. We did not attempt to deconvolve the traces as the spectra are too blended in our data set. The spectra presented here have been extracted using a wider aperture than all the other objects to ensure we included all the flux from both components. More observations of this object are needed to reliably determine the presence of a binary companion.

\subsubsection{2MASS J09240328+3653444 (L2+T3)}
\begin{figure*}
\centering
\includegraphics[scale=0.65]{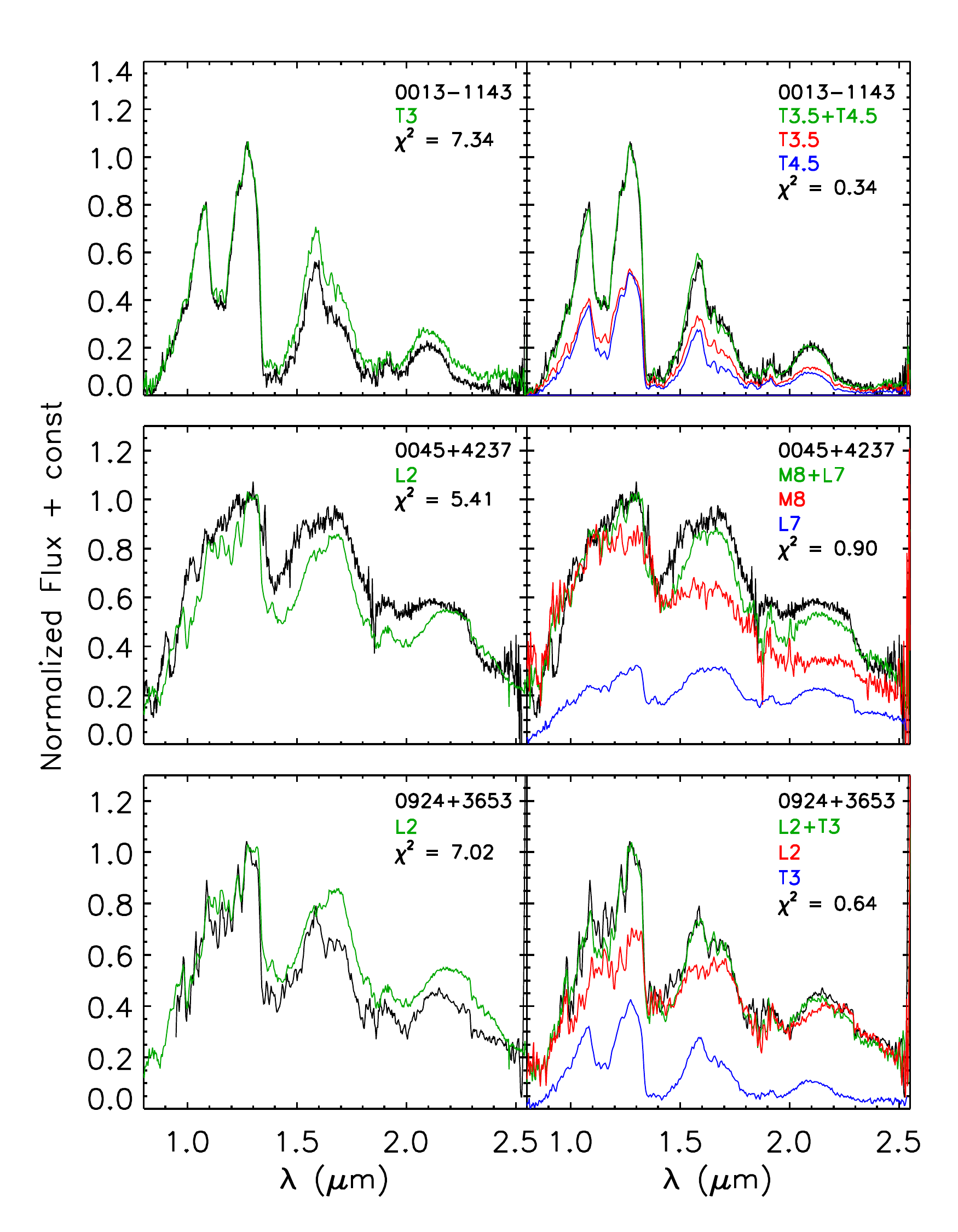}
\caption{Spectra of all objects identified as candidate unresolved binaries. The left panels show comparisons to spectra (in green) that fit the 0.95-1.35 \micron\ continuum best. The right panels show two-component templates (also in green) that fit best over 0.8-2.5 \micron; the individual component contributions are shown in red and blue and are scaled by their relative contributions. The quoted $\chi^2$ values are the smallest ones for, respectively, single- and binary-template fits over the entire 0.8-2.5 \micron\ range.  The comparison spectra from left to right and top to bottom are: T3 (2MASS J12095613$-$1004008; \citealp{burgasser04}), T3.5 (SDSSp J175032.96+175903.9; \citealp{burgasser04}) and T4.5 (2MASS J05591914$-$1404488; \citealp{burgasser06}); L2 (2MASS J13054019$-$2541059; \citealp{burgasser07}), M8 (2MASS J02481204+2445141; \citealp{kirkpatrick10}) and L7 (SDSS J140023.12+433822.3; \citealp{burgasser10}); L2 (2MASS J13054019$-$2541059; \citealp{burgasser07}), L2 (2MASS J12304602+2827515; \citealp{sheppard09}) and T3 (2MASS J12095613$-$1004008; \citealp{burgasser04}).}
\label{fig:bin1}
\end{figure*}

This object is a likely unresolved binary with an L2 primary component and a T3 secondary component. Figure \ref{fig:bin1} shows that an L2 object fits the $J$-band of 2M J0924+3653 relatively well but fails to properly match the CH$_4$ features in the $H$- and $K$-bands and the water feature between the $Y$- and $J$-bands. A composite template of an L2 and a T3 dwarf more closely matches the spectrum of this object. This object satisfies four of the twelve binary index selection criteria given in Table 4 of \cite{bardalez14} making it a weak binary candidate, according to that classification scheme.

\subsection{Normal Ultra-cool Dwarfs} \label{subsec:norm}

We classify 80 of our candidates as normal L dwarfs, i.e., they do not have any readily apparent peculiarities based on their comparison to SpeX spectral standards. We also identified 14 candidates as normal M dwarfs.  These were included in our program likely because the $i-z$ and $z-J$ colors of late-M dwarfs are close to the limits of our color selection criteria (Section~\ref{sec:selection}), and because they may have been subject to flux-overestimation bias at $J$-band (Section~3.3 of P1). The spectra of all these objects are included in Figure~\ref{fig:spectra}.

\subsection{False-Positive Ultra-cool Dwarfs} \label{subsec:fp}

\begin{figure*}
\centering
\includegraphics[scale=0.65]{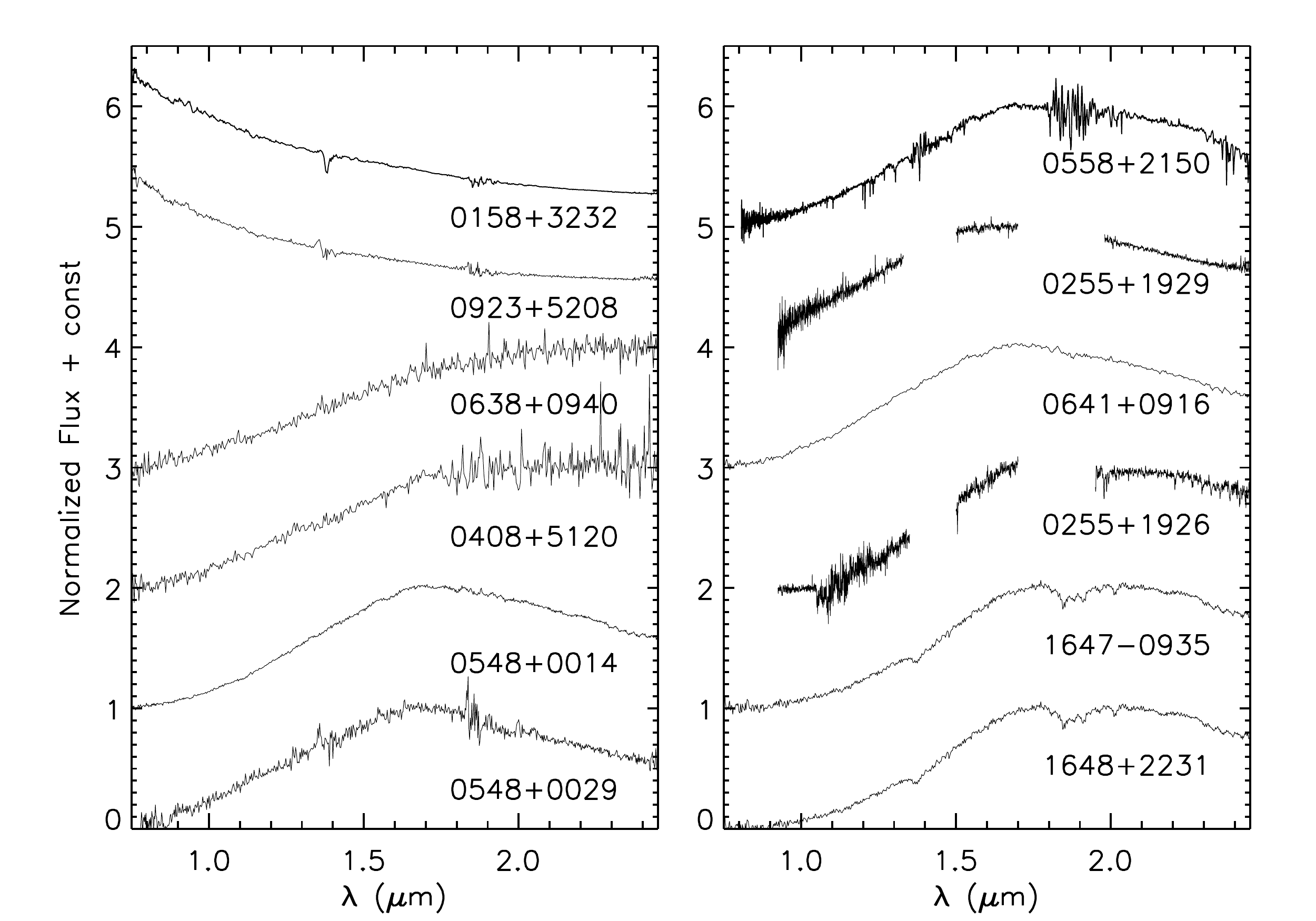}
\caption{SpeX prism, GNIRS cross-dispersed and Magellan/FIRE prism spectra of the 12 unknown objects that mistakenly passed our candidate selection criteria. }
\label{fig:fp}
\end{figure*}

We identified 5 false-positives in P1 and 7 in the concluding portion of the survey presented here. The spectra of these objects were taken with SpeX, GNIRS and FIRE (see P1 for instrument set up and extraction details for the Magellan/FIRE data). The spectra are presented in Figure \ref{fig:fp} and are grouped by spectral similarity. Their synthetic colors and photometric magnitudes are presented in Tables~\ref{tab:results} and \ref{tab:fp}, respectively. As can be seen in Figure \ref{fig:fp}, the spectra of these objects do not appear to be those of ultra-cool dwarfs. The first two of these objects in Figure \ref{fig:fp} look like they might be warmer stars and were most likely nearby stars that were targeted by mistake. We treat these objects as false-positives for the purpose of the survey statistics even though the candidates may indeed be ultra-cool dwarfs. All of the other objects have steep red slopes in the $z$- and $J$-bands, peak in the $H$-band, and are either relatively flat or taper off into the $K$-band. The $z-J$ and $J-K_s$ colors of these objects are similar to those of ultra-cool dwarfs and would have passed the selection criteria quite easily. We also notice a large discrepancy between the photometric and synthetic colors for these red objects (Tables~\ref{tab:results} and \ref{tab:fp}). This suggests that the photometric magnitudes for these objects are not accurate. We believe that all of the latter objects are extragalactic in origin.

%\clearpage
%\begin{turnpage}
\begin{deluxetable*}{cccccccccccc}
%\rotate
\tabletypesize{\scriptsize}
\tablecolumns{12}
\tablewidth{0pt}
\tablecaption{Unknown Object Properties
\label{tab:fp}}
\tablehead{
\colhead{2MASS ID} & \colhead{Telescope/} & \colhead{Survey\tablenotemark{a}} & \colhead{$z$} & \colhead{$J$} & \colhead{$H$} & \colhead{$K_s$} & \colhead{$W1$} & \colhead{$W2$} \\
\colhead{(J2000)} & \colhead{Instrument} & \colhead{Portion} & \colhead{(mag)} & \colhead{(mag)} & \colhead{(mag)} & \colhead{(mag)} & \colhead{(mag)} & \colhead{(mag)} }
\startdata
2MASS J01581172+3232013 & IRTF/SpeX & P1 & 18.80 $\pm$ 0.04 & 16.04 $\pm$ 0.07 & 14.89 $\pm$ 0.07 & 14.16 $\pm$ 0.05 & 13.69 $\pm$ 0.03 & 13.39 $\pm$ 0.03 \\
2MASS J02555058+1926476 & Gemini/GNIRS & P2 & 20.03 $\pm$ 0.10 & 17.33 $\pm$ 0.10 & 16.42 $\pm$ 0.24 & 15.20 $\pm$ 0.15 & 14.93 $\pm$ 0.04 & 14.54 $\pm$ 0.05 \\
2MASS J02553101+1929356 & Gemini/GNIRS & P2 & 20.40 $\pm$ 0.16 & 17.48 $\pm$ 0.28 & 15.34 $\pm$ 0.09 & 14.44 $\pm$ 0.08 & 14.30 $\pm$ 0.03 & 13.91 $\pm$ 0.04 \\
2MASS J04084337+5120524 & IRTF/SpeX & P1 & 19.48 $\pm$ 0.06 & 16.65 $\pm$ 0.15 & 14.64 $\pm$ 0.04 & 13.42 $\pm$ 0.04 & 12.09 $\pm$ 0.02 & 11.10 $\pm$ 0.02 \\
2MASS J05484895+0014367 & IRTF/SpeX & P2 & 19.21 $\pm$ 0.06 & 15.84 $\pm$ 0.08 & 13.98 $\pm$ 0.03 & 13.17 $\pm$ 0.03 & 12.87 $\pm$ 0.03 & 12.60 $\pm$ 0.03 \\
2MASS J05480405+0029264 & IRTF/SpeX & P2 & 20.11 $\pm$ 0.10 & 16.61 $\pm$ 0.15 & 15.00 $\pm$ 0.06 & 14.40 $\pm$ 0.08 & 14.03 $\pm$ 0.03 & 13.73 $\pm$ 0.04 \\
2MASS J05584262+2150121 & Magellan/FIRE & P1 & 19.19 $\pm$ 0.05 & 15.80 $\pm$ 0.07 & 14.00 $\pm$ 0.04 & 13.07 $\pm$ 0.02 & 12.53 $\pm$ 0.02 & 12.30 $\pm$ 0.03 \\
2MASS J06380876+0940084 & IRTF/SpeX & P1 & 20.44 $\pm$ 0.16 & 16.93 $\pm$ 0.20 & 15.08 $\pm$ 0.06 & 13.62 $\pm$ 0.04 & 12.24 $\pm$ 0.02 & 11.27 $\pm$ 0.02 \\
2MASS J06415196+0916111 & IRTF/SpeX & P1 & 19.63 $\pm$ 0.08 & 16.00 $\pm$ 0.09 & 14.18 $\pm$ 0.07 & 13.29 $\pm$ 0.05 & 12.65 $\pm$ 0.02 & 12.35 $\pm$ 0.03 \\
2MASS J09232257+5208598 & IRTF/SpeX & P2 & 19.13 $\pm$ 0.05 & 16.51 $\pm$ 0.12 & 16.38 $\pm$ 0.28 & 15.55 $\pm$ 0.20 & 15.32 $\pm$ 0.04 & 15.00 $\pm$ 0.07 \\
2MASS J16472470--0935294 & IRTF/SpeX & P2 & 20.44 $\pm$ 0.14 & 17.45 $\pm$ 0.30 & 15.57 $\pm$ 0.12 & 14.80 $\pm$ 0.10 & 14.49 $\pm$ 0.03 & 14.32 $\pm$ 0.05 \\
2MASS J16484099+2231397 & IRTF/SpeX & P2 & 19.18 $\pm$ 0.05 & 16.26 $\pm$ 0.09 & 15.20 $\pm$ 0.08 & 14.54 $\pm$ 0.09 & 13.91 $\pm$ 0.03 & 13.48 $\pm$ 0.03 \\
\enddata
%\tablecomments{}
\tablenotetext{a}{\footnotesize We indicate in which portion of the survey these objects were identified: P1 --- \cite{kellogg15}, P2 --- this work.}
\end{deluxetable*}
%\end{turnpage}
%\clearpage

\section{Population Statistics} \label{sec:stats}

We have completed a survey to identify unusual brown dwarfs in the SDSS and 2MASS catalogs. In the first portion of the survey (P1) we identified 4 peculiar ultra-cool dwarfs, 5 candidate L+T binaries, 17 normal L dwarfs, 13 normal M dwarfs, and one T dwarf candidate binary. In the concluding portion of the survey presented here, we have identified 7 additional peculiar ultra-cool dwarfs, 1 candidate L+T binary, 1 candidate M+L binary, 80 normal L dwarfs, 14 normal M dwarfs, and one more T dwarf candidate binary.  Table \ref{tab:results} summarizes the peculiarities of each object, as gleaned from analysis of their spectra. From the 144 new ultra-cool dwarfs discovered in the whole survey (including results from P1), we have identified 9 peculiarly red, 2 peculiarly blue, 7 candidate M+L and L+T dwarf binaries, two candidate T+T dwarf binaries, 97 normal L dwarfs, and 27 M dwarfs.

\begin{figure*}
\centering
\includegraphics[scale=0.65]{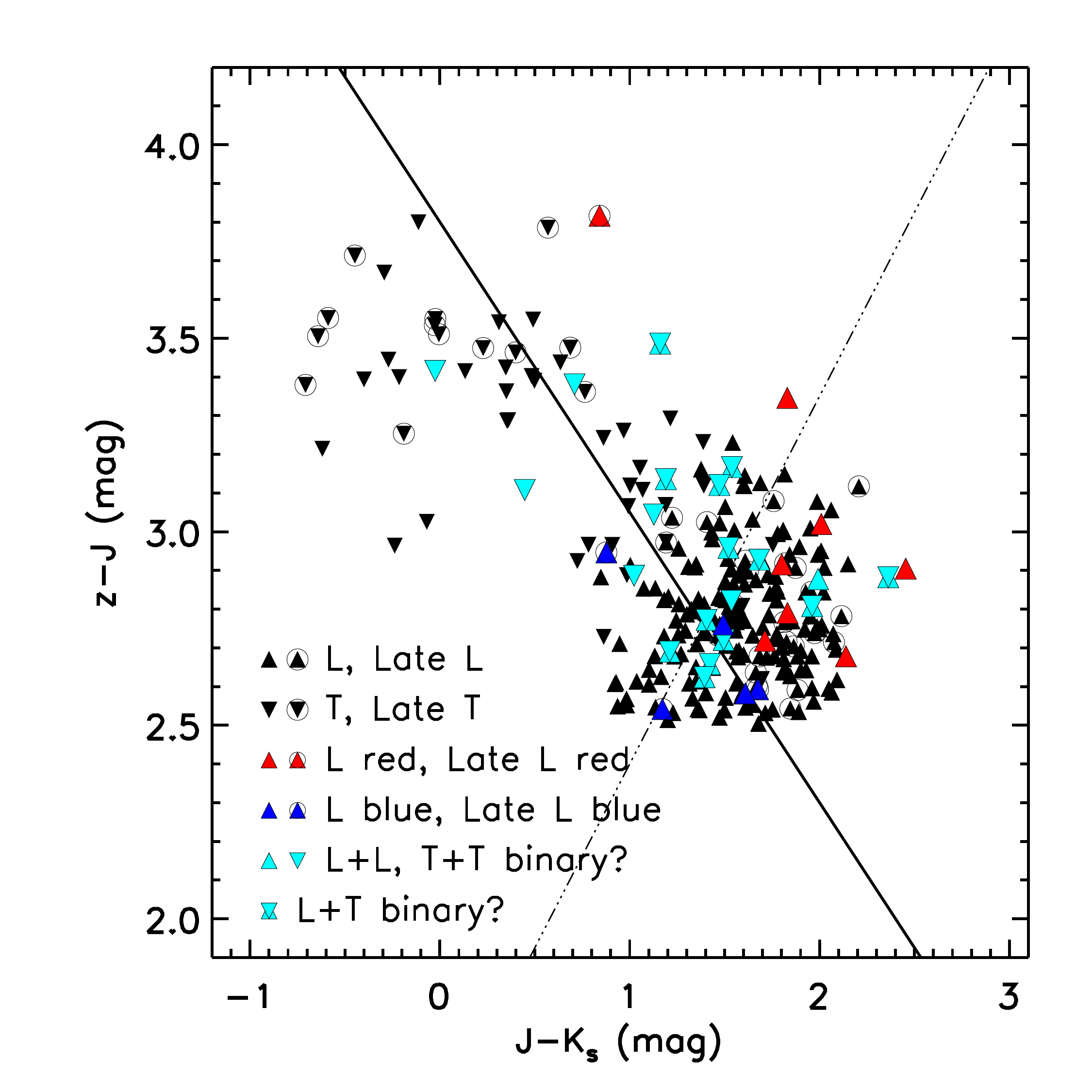}
\caption{Color-color diagram of all known L and T dwarfs recovered in our SDSS-2MASS-WISE cross-match. All symbols (upwards triangles - L dwarfs; downwards triangle - T dwarf) represent the photometric colors of the objects. The black symbols are ``normal" objects and the red and blue symbols are objects that have been identified as peculiar or binary. Late-L ($\geq$L7) and Late-T ($\geq$T6) dwarfs are denoted with circles. Our selection criteria from $\S$\ref{sec:selection} are denoted by the solid and dashed lines.}
\label{fig:rec}
\end{figure*}

Our goal is to assess the relative population of peculiar L and T dwarfs with $z-J >$ 2.5~mag, therefore, we also include the 276 previously known objects that our selection criteria recovered and use the classifications reported in the literature. There were additional ultra-cool dwarfs from SDSS reported in the J. Gagn\'e database\footnote{\url{https://jgagneastro.wordpress.com/list-of-ultracool-dwarfs/}} that did not pass our full set of selection criteria outlined in P1. Namely, these were objects detected in the shorter-wavelength bands (i.e. $r <$ 23~mag), had $z-J$ colors $<$ 2.5~mag, or had $H-W2$ colors $<$ 1.2~mag --- all typically L0 dwarfs. From the sample of previously known objects that we recovered, there are 8 unusually red objects, 5 unusually blue objects, and 21 candidate binary objects. There are also 186 normal L, 51 normal T, and 5 normal M dwarfs. All recovered objects are shown in Figure~\ref{fig:rec}. In this figure, we distinguish between late- and early-type objects where late-L type objects are $\geq$L7 and late-T objects are $\geq$T6. We also differentiate between binaries in which both components are L or T objects and binaries that have one component of each type. A summary of the types of objects found in our entire survey is presented in Table \ref{tab:stats}. Assessing the relative occurrence rates of the various peculiarities in our full sample of 420 ultra-cool dwarfs (144 new and 276 recovered) allows us to ascertain timescales of the associated phenomena.

\begin{deluxetable}{cccc}
\tabletypesize{\scriptsize}
\tablecolumns{4}
\tablewidth{0pt}
\tablecaption{Summary of Object Types
\label{tab:stats}}
\tablehead{
\colhead{Type} & \colhead{New} & \colhead{Recovered} & \colhead{Total} }
\startdata
M & 27 & 5 & 32 \\
L0--L6 & 88 & 163 & 251 \\
L7--L9 & 9 & 23 & 32 \\
T0--T5 & 0 & 38 & 38 \\
T6--T9 & 0 & 13 & 13 \\
Young L & 4 & 5 & 9 \\
Red L & 5 & 3 & 8 \\
Blue L & 2 & 5 & 7 \\
L+L binary & 0 & 3 & 3 \\
T+T binary & 2 & 7 & 9 \\
L+T binary & 6 & 11 & 17 \\
M+L binary & 1 & 0 & 1 \\
Total & 144 & 276 & 420 \\
%\hline
%\hline
%Satisfied red criterion alone & 62 & 126 & 188 \\
%Satisfied binary criterion alone & 21 & 66 & 87 \\
%Satisfied both criteria & 30 & 56 & 86 \\
%General ultra-cool dwarf candidates & 31 & 28 & 59 \\
\enddata
%\tablecomments{}
\end{deluxetable}

\subsection{Efficacy of the Survey} \label{subsec:efficacy}

We had a false positive rate of 11\% from the first portion of our survey (5 unknown objects out of 45 total candidates). We refined our visual selection for the second portion based on these false positives and reduced our rate to 6.2\% (7 unknown objects out of 113 total remaining candidates). We exclude these objects when discussing the statistics of our full survey and only consider the 144 confirmed new and 276 previously known ultra-cool dwarfs.

From the total 420 objects that our color criteria selected and that we confirmed to be ultra-cool dwarfs, 17 (4.0\%) are peculiarly red, 7 (1.7\%) are peculiarly blue and 30 (7.1\%) are candidate binaries. The number of peculiarly red objects in our sample is statistically equivalent with that of the 4.6\% of red objects found in an unbiased sample of L and T dwarfs from \cite{faherty09}. Our peculiarly red selection criterion, however, identified 9 new red objects among the sample of 92 candidates (9.8\%). None are peculiarly red out of the 31 general ultra-cool dwarf candidate sample so our selection technique successfully identified all of the peculiarly red objects in the sample. Including the previously known ultra-cool dwarfs, 6.2\% of objects were red among the sample of objects that satisfied the peculiarly red criterion (17 of 274). 

We had a better success rate for objects that satisfied the candidate binary criterion. Of the 51 new binary candidates, 10 (20\%) were either peculiarly blue (2) or are potential binaries (8). Including the previously known ultra-cool dwarfs, 13\% were unusually blue (3) or candidate binaries (20) among the sample of objects that satisfied the candidate binary criterion (173).

When we compare our newly discovered peculiarly red brown dwarfs (9) to the number of red L and T dwarfs that were already known and were recovered with our selection criterion (8), we see that we increased the sample in this color space by a factor of 2. Similarly, we see that we increased the sample of candidate binaries by a factor of 1.4. With only 2 new unusually blue discoveries, we did not significantly impact the statistics of these types of objects. This is unsurprising as we were not targeting these objects. 

We also did not uncover many L+L or T+T binaries. Binaries where both components have similar spectral morphologies are harder to discern from single low to moderate resolution spectra so are more difficult to identify. An example of this is the planetary-mass object, 2MASS J11193254--1137466, that we discovered in the first portion of the survey (L7; P1) that was later resolved into a binary system (L7+L7; \citealp{best17b}). We identified this object as peculiar based on its extremely red near-infrared colors and weak alkali absorption features but did not suspect unresolved binarity from our spectra.

\subsection{Unusually Red Objects} \label{subsec:redstats}

Among the 17 unusually red objects from the whole survey of 420 ultra-cool dwarfs, 9 (2.1\%) are young and 8 (1.9\%) are red with no signatures of youth. The $\sim$2\% fraction of low-gravity ultra-cool L dwarfs is consistent with a $\lesssim$200~Myr age for these objects under the assumption of a constant star-formation history for the Milky Way (e.g. \citealp{burrows93,marley96}). This is also consistent with evolutionary models such as those in \cite{burrows01} and \cite{baraffe15} that say the radii of ultra-cool dwarfs becomes constant after $\sim$200~Myr and studies of objects in young stellar associations such as \cite{allers13} and \cite{liu16} which show that associations older than $\sim$200~Myr have a much smaller fraction of objects that show signatures of youth than younger associations.

For the 1.9\% of objects that are unusually red with no signatures of youth, there has thus far not been a satisfactory explanation of their redness. We suggest that these objects may have reached a point where their surface gravities are not low enough to have distinguishing alkali line strengths from field objects in low resolution spectra, but are still young enough that their dust has not completely settled. This dust could be the sub-micron particles that potentially play a significant role in the reddening of ultra-cool dwarf spectra and settle less efficiently than the 1-100 \micron\ grains \citep{hiranaka16}. If this is the case, then the peculiarly red color that is detectable in their optical to near-infrared SEDs is a better indicator of moderate youth than individual absorption features. Another possible explanation is that these objects were formed in environments with higher metallicity. If this is the case, then $\sim$2\% of objects within the sensitivity limit of the SDSS, 2MASS and WISE surveys were born in these conditions.

\cite{kirkpatrick10} offer a further speculation on the nature of these unusually red objects. They find that the space velocities of these objects are similar to those of unusually blue L dwarfs that do not have any signatures of low-metallicity. They suggest that these two kinds of objects could be related and their spectral appearance could be the result of different viewing angles. If clouds are not homogeneously distributed in latitude or if cloud properties such as grain size and thickness vary in latitude, then viewing an object pole-on versus equator-on would change the spectral morphology. However, if this were the case, then we would expect the number of objects with these properties to be higher because such conditions would be ubiquitous for all brown dwarfs.

\subsection{Candidate Binaries} \label{subsec:binstats}

Our binary selection criterion was designed to identify L+T binaries. In these objects, both components contribute equally to the flux in the $J$-band but unequally in the $z$- and $K$-bands, making the $z-J$ colors red and the $J-K_s$ colors blue. We identified and recovered 30 objects that are candidate binaries out of the 420 objects in the full survey, 17 of which are binaries with one L and one T component. We did not find any new L+L binary candidates and only two new T+T binary candidates: likely because they are difficult to identify even from their spectra as the two components would be more similar in spectral morphology than the components in an L+T binary. These objects would look more like single objects and could only be identified as binaries through other means.

Our survey also recovered 51 known single T dwarfs and 9 candidate T+T binaries, 2 of which were new.  These represent the entire population of T dwarfs in SDSS.  The vast majority were already known from previous searches for T dwarfs in SDSS. Both of the newly discovered objects we categorized as candidate binaries, although they could also be highly variable T dwarfs. 

\subsection{WISE Colors} \label{subsec:wise}
\begin{figure*}[h]
\centering
\includegraphics[scale=0.6]{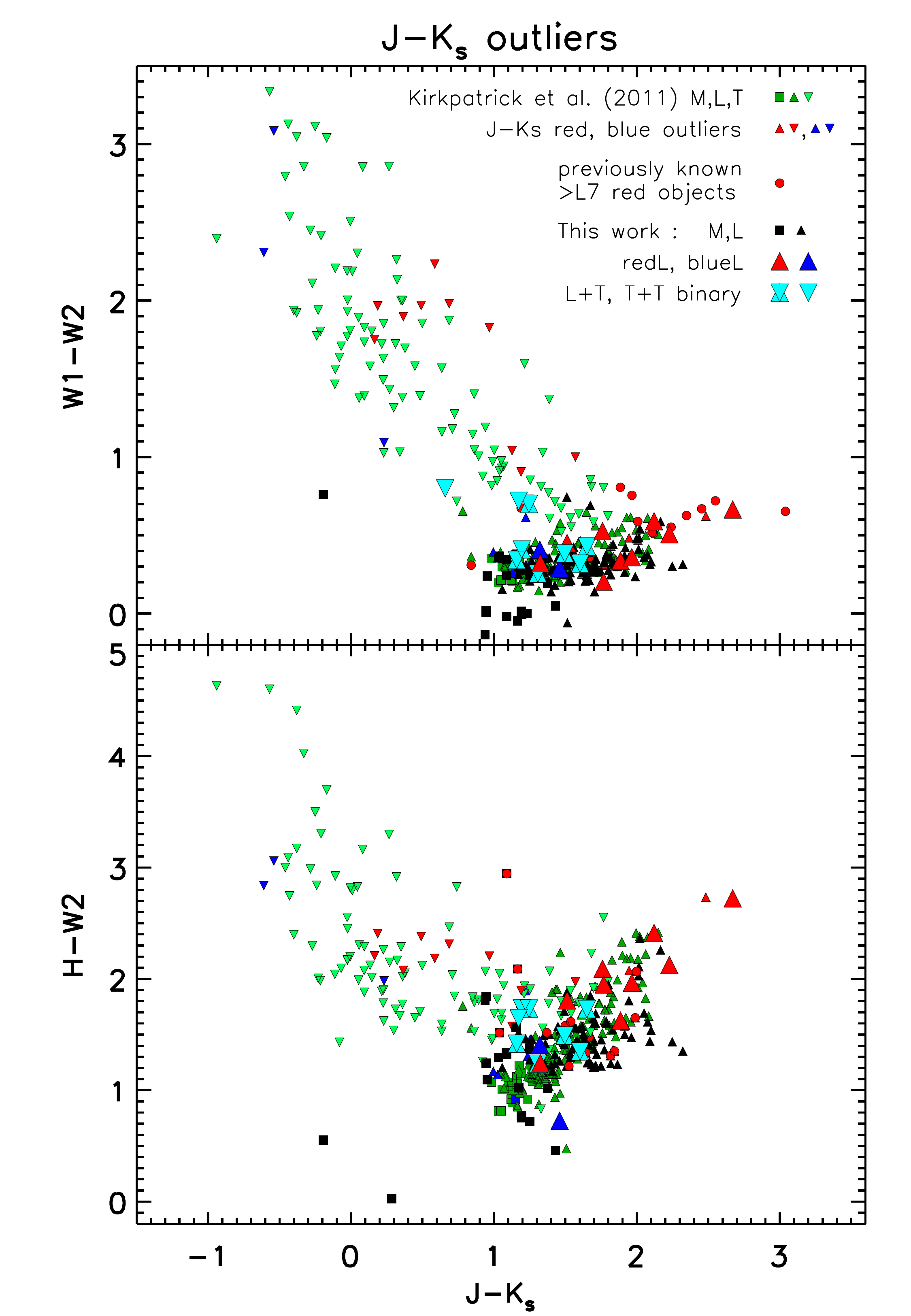}
\caption{Photometric color-color diagrams of objects from \cite{kirkpatrick11}. Upwards and downwards triangles denote L and T dwarfs, respectively.  Red symbols denote objects with $J-K_s$ colors $>$2$\sigma$ redder than the mean for their spectral type \cite{faherty09,faherty13}.  Blue symbols denote objects that are $>$2$\sigma$ bluer. Large symbols represent peculiar objects identified in this work and normal symbols represent the normal objects. Red circles indicate other previously known red ultra-cool dwarfs not in \cite{kirkpatrick11} with spectral types of L7 and later.}
\label{fig:jkout}
\end{figure*}

As done in P1, we investigated whether the $J-K_s$ color outliers also have unusual colors in the $W1$ and $W2$ WISE bands. We have compared the objects that we discovered in this survey to known ultra-cool dwarfs in the AllWISE catalog published in \cite{kirkpatrick11}. We determined the red and blue outliers from this data set using the median $J-K_s$ colors and standard deviations for each spectral type from \cite{faherty09} and \cite{faherty13}. We confirm the results of P1 that L dwarfs with the very reddest $J-K_s$ colors are clearly distinguishable from the locus of L dwarfs on a $J-K_{s}$ vs.\ $H-W2$ and $J-K_{s}$ vs.\ $W1-W2$ diagram (Fig. \ref{fig:jkout}) mainly because of their red $J-K_s$ colors. T dwarfs with peculiarly red $J-K_s$ colors are slightly redder in both $H-W2$ and $W1-W2$, and the peculiarly blue L or T dwarfs and candidate binaries are not distinguishable from the normal population. 

\section{Conclusions} \label{sec:concl}

We have completed a survey to identify ultra-cool dwarfs with peculiar photometric colors in the SDSS, 2MASS and WISE catalogs. In the concluding portion of our survey, we have found two new candidate very low-gravity, planetary-mass objects: 2MASS J00133470+1109403 (L1) and 2MASS J00440332+0228112 (L7).  The latter was independently found to be a planetary-mass object with a high probability of $\beta$ Pic membership \citep{schneider17}. Our survey also identified one of the reddest objects with no signatures of youth known to date: the L6 dwarf 2MASS J03530419+0418193. A detailed study of this object may give clues as to the nature of its extreme red color as no satisfactory answer has been found to explain such objects so far. 

With spectroscopic observations of the candidates from the first portion of the survey (P1) and from this work, we confirmed that 20 of our new 144 ultra-cool dwarfs are unusually red, unusually blue, or are candidate binaries. Including the 276 previously known objects that we recovered with our selection criteria and the 144 objects discovered in this survey (420 objects total), 4.0\% (17) are unusually red, 1.7\% (7) are unusually blue and 7.1\% (30) are candidate binaries. We find that there are roughly as many L+T binaries in our sample as binaries of any other kind combined, likely because L+L or T+T binaries would be difficult to identify from low to moderate resolution spectra alone. 

We also find that there are almost equal numbers of red L dwarfs that are young based on weak potassium absorption strengths (2.1\%, 9/420) and red L dwarfs with normal potassium absorption (1.9\%, 8/420).  The first population are likely younger than 200 Myr: the approximate age where contraction mostly halts in $\gtrsim$13 \MJup\ brown dwarfs.  The latter population may be only slightly older --- by up to a factor of 2 --- or may alternatively be more metal-rich.

\acknowledgements
This work was supported by the NASA Astrophysical Data Analysis Program through award No.\ NNX11AB18G to S.M. at Stony Brook University and by an NSERC Discovery grant to S.M. at The University of Western Ontario. This research has made use of data obtained from or software provided by the US Virtual Astronomical Observatory, which is sponsored by the National Science Foundation and the National Aeronautics and Space Administration. Based on observations as part of Programs GN-2015A-Q-57, GN-2015B-Q-79 and GN-2017A-Q-44 obtained at the Gemini Observatory (processed using the Gemini IRAF package), which is operated by the Association of Universities for Research in Astronomy, Inc., under a cooperative agreement with the NSF on behalf of the Gemini partnership: the National Science Foundation (United States), the National Research Council (Canada), CONICYT (Chile), Ministerio de Ciencia, Tecnolog\'{i}a e Innovaci\'{o}n Productiva (Argentina), and Minist\'{e}rio da Ci\^{e}ncia, Tecnologia e Inova\c{c}\~{a}o (Brazil). The authors wish to recognize and acknowledge the very significant cultural role and reverence that the summit of Mauna Kea has always had within the indigenous Hawaiian community.  We are most fortunate to have the opportunity to conduct observations from this mountain.  This publication makes use of data products from the Two Micron All Sky Survey, which is a joint project of the University of Massachusetts and the Infrared Processing and Analysis Center/California Institute of Technology, funded by the National Aeronautics and Space Administration and the National Science Foundation. Funding for SDSS-III has been provided by the Alfred P.\ Sloan Foundation, the Participating Institutions, the National Science Foundation, and the U.S. Department of Energy Office of Science. The SDSS-III web site is http://www.sdss3.org/.  SDSS-III is managed by the Astrophysical Research Consortium for the Participating Institutions of the SDSS-III Collaboration including the University of Arizona, the Brazilian Participation Group, Brookhaven National Laboratory, Carnegie Mellon University, University of Florida, the French Participation Group, the German Participation Group, Harvard University, the Instituto de Astrofisica de Canarias, the Michigan State/Notre Dame/JINA Participation Group, Johns Hopkins University, Lawrence Berkeley National Laboratory, Max Planck Institute for Astrophysics, Max Planck Institute for Extraterrestrial Physics, New Mexico State University, New York University, Ohio State University, Pennsylvania State University, University of Portsmouth, Princeton University, the Spanish Participation Group, University of Tokyo, University of Utah, Vanderbilt University, University of Virginia, University of Washington, and Yale University. This publication makes use of data products from the Wide-field Infrared Survey Explorer, which is a joint project of the University of California, Los Angeles, and the Jet Propulsion Laboratory/California Institute of Technology, funded by the National Aeronautics and Space Administration. 

\facilities{IRTF (SpeX), Gemini North (GNIRS)}
\software{IDL, IRAF}
\facilities{SDSS, 2MASS, WISE}

%\bibliography{bibliography.bib}
\bibliographystyle{apj}

\end{document}